\documentclass[prd,superscriptaddress,onecolumn,showpacs,11pt,msmath,preprintnumbers,showkeys]{revtex4}
\usepackage{wrapfig,rotating}
\usepackage{graphicx,epsfig,color}

\makeatletter

\newenvironment{figurehere}
{\def\@captype{figure}}
{}
\makeatother

\usepackage{graphicx}
\usepackage{dcolumn}
\usepackage{bm}

\def\beq{\begin{equation}}
\def\eeq{\end{equation}}
\def\beeq{\begin{eqnarray}}
\def\eeeq{\end{eqnarray}}

\def\as{\alpha_{\mbox{\rm\scriptsize s}}}

\def\cO#1{{\cal{O}}\left(#1\right)}
\newcommand \Pomeron {I\!\!P}
\newcommand \R {I\!\!R}

\def\effs {$\sigma_{\textrm{\tiny eff}}\,$}
\def\2GPD{$_2\mbox{GPD}$}
\def\GeV{{\rm Ge\!V}}
\def\oo{$1 \otimes 1$}
\def\12{$1\otimes 2$}
\def\22{$2 \otimes 2$}
\def\eff{{\mbox{\scriptsize eff}}}
\def\half{\frac{1}{2}}

\def\Qsep{Q_{\mbox{\rm\scriptsize sep}}}
\def\Qsep2{Q^2_{\mbox{\rm\scriptsize sep}}}
\def\beeq{\begin{eqnarray}}
\def\eeeq{\end{eqnarray}}
\newcommand{\be}{\begin{equation}}
\newcommand{\ee}{\end{equation}}
\newcommand{\bea}{\begin{eqnarray}}
\newcommand{\eea}{\end{eqnarray}}

\begin{document}
 \title{Multiparton pp and pA collisions -- from geometry to parton-- parton correlations$\dagger$}
 \author{B.\ Blok$^{1}$,
M. Strikman$^2$ \\[2mm] \normalsize $^1$ Department of Physics, Technion -- Israel Institute of Technology,
 Haifa, Israel\\
 \normalsize $^2$Physics Department, Pennsylvania State University, University Park,USA }
 \begin{abstract}
We derive expressions for  the cross section of the multiparton interactions based on the analysis of the relevant Feynman diagrams.  We express the cross sections  through
 the double (triple, ...)  generalized parton distributions (GPDs).  In the mean field approximation for the double GPDs  the answer is expressed through the integral over two gluon form factor which was
  measured in   the
 exclusive DIS vector meson production.
 We explain under what conditions the  derived expressions  correspond to an intuitive picture of hard interactions in the impact parameter representation.
The mean field approximation in which correlations of the partons are neglected fail to explain the data, while pQCD induced correlation enhance large $p_\perp$ and $ 0.001 < x < 0.1$ typically enhance the cross section by a factor of 1.5 -- 2 explaining the current data. We argue that in the small x kinematics ($10^{-4} \le x \le 10^{-3}$) where effects of perturbative correlations diminish,  the  nonperturbative mechanism kicks in and  generates positive correlations comparable in
magnitude with the perturbative ones. We explain how our technique can be used for calculations of MPI in the proton - nucleus scattering.
The interplay of hard interactions and underlying event is  discussed, as well as different geometric pictures for each of MPI mechanisms-pQCD, nonperturbative correlations and mean field.
Predictions for value of \effs  for various processes and a wide range of kinematics are given.
 We show that
together different MPI mechanisms give good description of experimental data, both at Tvatron, and LHC, including the central kinematics studied by ATLAS and CMS detectors, and forward (heavy flavors) kinematics studied by LHCb.
\vspace{1cm}

\noindent
$\dagger$ To be published in "Multiple parton Interactions at the
LHC", P. Bartalini and J. Gaunt eds, World Scientific
 \end{abstract}

   \maketitle
 \thispagestyle{empty}

 \vfill

 \section{Introduction}
 \par It is widely realized now that hard {\em Multiple Parton Interactions}\/ (MPI)
occur with a probability of the order one in  typical inelastic LHC proton-proton $pp$ collisions. Indeed the ratio of the  integral of the inclusive jet cross section with transverse momenta $  p_{\perp} \ge  \mbox{few GeV}$ and  $ \sigma_{inel}(NN) $ gives the average multiplicity of hard collisions (dijet production) larger than one, see e.g. \cite{Sjostrand,Jung}.
Hence   MPI play an important role in the description of inelastic  $pp$ collisions.
MPI were first introduced in the eighties
 \cite{TreleaniPaver82,mufti} and in the last decade became a subject of a number of the theoretical studies,  see e.g. \cite{stirling,BDFS1,Diehl,stirling1,BDFS2,Diehl2,BDFS3,BDFS4,Manohar1,Manohar2,Gauntnew,BO,Gauntdiehl,Gauntnew1,mpi2014,mpi2015,BG1,BG2,BS1,ST,BS2,BS3,BS4,BS5} and references therein.
\par Also,
in the past several years  a number of Double Parton Scattering (DPS) measurements in different channels    were carried out
\cite{Tevatron1,Tevatron2,Tevatron3,cms1,atlas,Aad:2013bjm,cms2,Belyaev,LHCb,LHCb1,LHCb2}, while many Monte Carlo (MC)  event generators now incorporate MPIs
\cite{Butter,Skands,gieseke1,gieseke2,gieseke3,Sjodmok,Sjostrand:2014zea,Corke:2011yy,Corke:2010yf,Herwig,Lund}.
\par The double parton scattering (DPS) cross section is traditionally parameterized  as
 \beq
{d\sigma(4\to 4) \over d\Omega_1 d \Omega_2} = {1\over \sigma_{eff}}
{d\sigma(2\to 2) \over d\Omega_1 }{d\sigma(2\to 2) \over d\Omega_2},
\label{sigmaeff}
\eeq
where $\Omega_i$ is the phase volume for production of a pair of jets where $\sigma_{eff}$ is a priori a function of $x_i, p_{t_i}$
  Initially it was conjectured \cite{TreleaniPaver82} that parameter $\sigma_{eff}$ is related to the total inelastic cross  section of the hadron - hadron interactions.

\begin{figure}[t]
\hspace{1.7cm}\includegraphics[width=.70\textwidth]{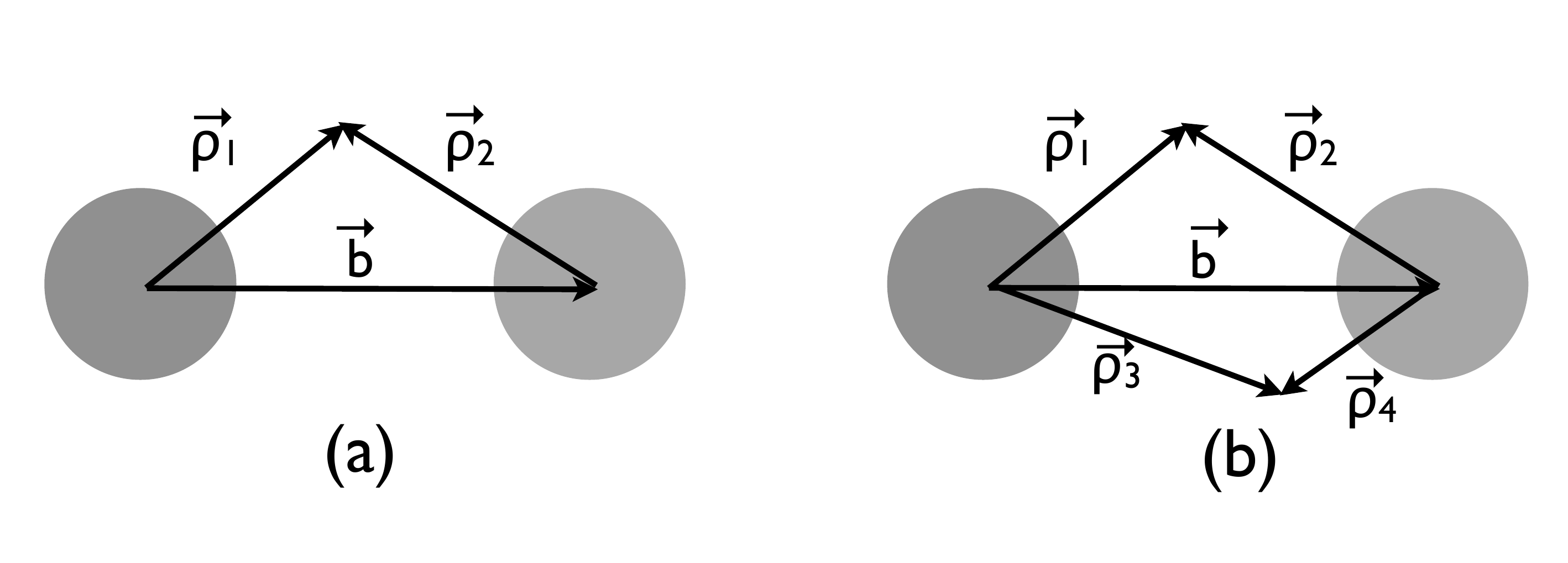}
\caption[]{Geometry of one and two hard collisions in impact parameter picture. }
\label{overlap4}
\end{figure}

   Later on within the framework of the geometric picture implemented in the Monte Carlo models \effs was written as a convolution of the four single parton impact parameter distributions,
  $g(\rho_i)$ assuming that these distributions do not depend on $x$ and on flavor, cf. Fig.~\ref{overlap4}.
  \beq
  {1\over \sigma_{eff}}= \int d^2 \rho_i d^2b g(\rho_1) g(\rho_3)g(\rho_2) g(\rho_3)g(\rho_4)
  \delta (\vec{\rho}_1 - \vec{\rho}_3 -\vec{b})
  \delta (\vec{\rho}_2 - \vec{\rho}_4 -\vec{b}).
  \label{geom}
  \eeq
  One can see from Eq.\ref{geom}
  that the factor \effs characterizes the transverse area occupied by the
partons participating in  two hard collisions. It also includes effect  of possible longitudinal correlations between the partons.

  Parameters of this distribution were chosen to reproduce the MPI data obtained at the Tevatron which reported  \effs $\approx$ 15 mb.

  Further study used the QCD factorization theorem for the exclusive  vector meson production to extract $g(\rho, x|Q^2)$ from the photo/electro production data. Under assumption that partons in colliding nucleons are not correlated a much larger \effs $\ge 30$ mb was found \cite{FSW}. This strongly suggested that significant    parton - parton correlations are present  in nucleons.

  In this paper we will summarize our studies of  the mechanisms which generate perturbative and nonperturbative correlations between the partons and allow to explain many features of the data.
  In particular we explain the geometry of MPI and show that the MPI cross section is given by the  sum of the mean field contribution,
  pQCD and nonperturbative mechanisms, connected with nonfactorizable initial conditions.   Each of these three mechanisms
  corresponds to its different range of impact parameters, (with the mean field one being most central).  Together  they lead to a good agreement of experimental MPI cross sections.

    The text is organized as following.

  In sec. 2 we present the geometrical picture of MPI and explain that hard collisions, in average correspond to much smaller impact parameters than th minimum bias inelastic collisions.
  In sec. 3 we review the parton level calculation of the DPS using Feynman diagram analysis
  which allows to express the DPS cross section through the convolution of two  double generalized parton distributions (GPD).  The  double GPDs in the mean filed approximation are  expressed through a product of single  GPDs which are extracted from the studies of the exclusive vector meson production.
    In section 4 we analyze contribution to the DPS of the correlation mechanism induced  by the pQCD evolution. General expressions are derived both for the cross section differential in jet imbalances $\delta_{ij}$
    and the cross section integrated over
 $\delta_{ij}$.

 The  numerical results for the contribution of pQCD correlation mechanism are  presented in sec. 5. We find that  perturbative  mechanism may enhance the DPS rates at large $p_{\perp}$ (large virtualities) and $x \sim 10^{-2} \div 10^{-3}$ by a factor 1.5 -- 2 allowing to explain the observed rates for a number of DPS processes.

 In section 6 we argue that a new soft mechanism of the parton - parton correlations  becomes important  for $ x \leq 10^{-3} $ which is due to presence of multiPomeron exchanges. We explain that this  mechanism is relevant for the
    for the understanding of the rate of minijet production  and as well as  the production of two D-mesons  in the forward kinematics studied by LHCb \cite{Belyaev,LHCb,LHCb1,LHCb2}.
In section 7 we apply our technique to calculate the rate of MPI in proton - nucleus collisions taking into account pQCD corrections to the parton model approximation \cite{ST} and finding that pQCD corrections further increase the ratio of MPI in pp and pA scattering \cite{BS2}.

    In section  8 we consider several consequences of the  different impact parameter localization of the minimum  bias and hard collisions. In particular we explain that b-space unitarity leads to requirement that jet production cross section should be suppressed was compared to the pQCD result even at large impact parameters.

    Our conclusions are presented in section 9.

 \section{Transverse picture of multiparton interactions}
 \label{transverse}

 \subsection{Impact parameter distribution in hard collisions}

 A natural framework for visualization of the MPI  is the impact parameter representation  of the collision. Indeed, in the high energy limit the angular momentum  conservation implies that the impact parameter $b$  becomes a good quantum number.    Also  the hard collisions are localized in the transverse plane at the relative distances $\sim 1/Q$ where $Q$ is transverse momentum transfer. Combined, they lead to an  intuitive picture of the MPI.

 To describe the transverse geometry of the $pp$ collisions with production of a dijet it is convenient to consider  probability to find a parton with given $x$ and transverse distance $\vec{\rho}$ from the nucleon transverse center of mass, $f_i(x_i,\vec{\rho}_i)$. This quantity allows a formal operator definition, and it is referred to as  the diagonal generalized parton distribution(GPD). It is related to non-diagonal GPDs which enter in the description of the exclusive meson production (see Appendix for discussion of the information on $\rho$ dependence of GPDs which is available from the studies of the exclusive vector meson production in the DIS.).

The inclusive cross section in the LT  pQCD regime does not depend on the  transverse structure of the colliding hadrons  - the cross section is expressed through the convolution of parton densities. Indeed, we can write
 \bea
 \sigma_h \propto \int d^2bd^2\rho_1d^2\rho_2 \delta(\rho_1+b-\rho_2)
 f_1(x_1,\rho_1)f_2(x_2,\rho_2)\sigma_{2\to 2} =\nonumber \\
  \int d^2bd^2\rho_1d^2\rho_2f_1(x_1,\rho_1)f_2(x_2,\rho_2)\sigma_{2\to 2}=f_1(x_1)f_2(x_2)\sigma_{2\to 2}.& &
 \eea
 Here at the last step we used the relation between diagonal GPD and PDF:
 $\int d^2\rho f_j(x,\rho,Q^2)= f_j(x,Q^2)$.

 At the same time,  as soon as one wants to describe  the structure of the final state in production of say dijets, it is important to know whether a hard process occurs at   different average impact parameters  than in  the
  minimum bias interactions. It turns out that  at the LHC energies a
 dijet trigger selects, in average,  a factor of two smaller impact parameters than in the minimum bias events.
 This implies   that the multijet activity, energy flow  should be much stronger in these events than in the minimum bias events.
 Obviously,  the
magnitude of the enhancement does depend on the transverse distribution of partons and on the correlation between the partons in the transverse plane.
 This information becomes available now. It is summarized in the Appendix.

 In the case of collisions with $N$ hard subprocesses the interaction picture corresponds to a pairwise  localization of $N $ partons of each of the nucleons at short distances (Fig.~\ref{overlap4}b), leading to the cross section of collision of hadrons $a$ and $b$ proportional to
 \beq
 \sigma_h^{(N)} \propto \int d^2b\prod_{i=1}^{i=N}
 d \rho_i d \rho_i^\prime  \delta(\rho_i+b-\rho_i^\prime )f_a(\rho_i, Q_i) f_b(\rho_i^{\prime}, Q_i).
 \label{generic}
 \eeq
 The geometric pairwise overlap  with $N$ partons of hadrons $a$ and $b$ nearby pairwise  provides a geometric factor $L^{N-1}$  in the cross section for $N$ hard collisions,
 where $L $ is the  linear scale  proportional to the transverse  linear scale of the colliding hadrons.
 Eq. \ref{generic} includes correlations between partons both on the hadronic distance scale and local correlations due to the QCD evolution. In the case of perturbative correlations when two partons of one of the colliding nucleons are close together the overlap factor is enhanced as compared to the uncorrelated case, see discussion in sec. 3.

 Using the information on the transverse spatial distribution
of partons in the nucleon, one can obtain the distribution over
impact parameters in $pp$ collisions with hard parton--parton
processes \cite{FSW}.  It is given by the overlap of two parton wave functions as depicted in Fig.~\ref{overlap4}.

The probability distribution of
$pp$ impact parameters in events with a given hard process, $P_2 (x_1, x_2, b|Q^2)$,   is given by the ratio of the cross section at given $b$ and the cross section integrated over $b$.
As a result
\begin{eqnarray}
P_2 (x_1, x_2, b|Q^2) &\equiv&
\int \! d^2\rho_1 \int \! d^2\rho_2 \;
\delta^{(2)} (\bm{b} - \bm{\rho}_1 + \bm{\rho}_2 )\nonumber\\[10pt]
&\times& F_{2g} (x_1, \rho_1 |Q^2 ) \; F_{2g} (x_2, \rho_2 |Q^2) \, ,
\label{P_2_def}
\end{eqnarray}
which obviously satisfies the  normalization condition
\beq
\int d^2b \, P_2 (x_1, x_2, b |Q^2) \;\; = \;\; 1.
\eeq
This distribution represents an essential tool for  phenomenological
studies of the underlying event in $pp$
collisions \cite{FSW,Frankfurt:2010ea}, see discussion in Sec. 7.

For the two  parametrizations of Eq.~(\ref{twogl_exp_dip}),
 Eq.~(\ref{P_2_def}) leads to  ( for $x\equiv x_1 = x_2$)
 \be
P_2 (x, b| Q^2) \; = \; \left\{
\begin{array}{l}
\displaystyle
(4\pi B_g)^{-1} \, \exp [-b^2/(4 B_g)] ,
\\[2ex]
\displaystyle
[m_g^2 /(12\pi)] \, (m_g b/2)^3 \, K_3 (m_{g} b) ,
\label{P_2_exp_dip}
\end{array}
\right.
\ee
where the parameters $B_g$ and $m_g$ are taken at the appropriate
values of $x$ and $Q^2$. Since $B_g$ increases with a decrease of $x$ , distribution over $b$ depends on $x$'s of the colliding partons and their virtualities, however this effect is pretty small for production of jets at central rapidities, see e.g. Figs.~4, 5 in \cite{Frankfurt:2010ea}.

{\it Comment} A word of caution is necessary here. The transverse distance $b$ for dijet events is defined as the distance between the transverse centers of mass of two nucleons. It may not coincide with  $b$ defined for soft interactions where soft partons play an important role. For example, if we consider dijet production due to the interaction of two partons with $x\sim 1$, $\rho_1,\rho_2 \sim 0$ since the transverse center of mass coincides with transverse position of the leading quark in the $x\to 1$ limit.
  As a result, $b$ for the hard collision will be close to zero. On the other hand the rest of the partons may interact in this case at th every different transverse coordinates. As a result,  such configurations may    contribute to the inelastic $pp$  cross section at much larger  $b$ for the soft interactions. However for the parton collisions at  $x_1,x_2 \ll 1$
the recoil effects  are small and so two values of $b$ should be close.

\subsection{Impact parameter distribution in minimum bias collisions}

The derived distribution should be compared to the distribution of the minimum bias inelastic collisions which could be expressed through  $\Gamma (s, b)$ that  is the profile function of the $pp$ elastic
amplitude ($\Gamma (s, b)=1$ if the interaction is completely absorptive at given $b$)
\be
P_{\mbox{in}} (s, b) \;\; = \;\;
\left[ 1 - |1 - \Gamma (s, b)|^2\right]\,  / \sigma_{\mbox{in}}(s) ,
\label{P_in_def}
\ee
where
$\int d^2 b \, P_{\mbox{in}} (s, b) = 1$.

\begin{figure}[t]
\hspace{1.7cm}\includegraphics[scale=0.8]{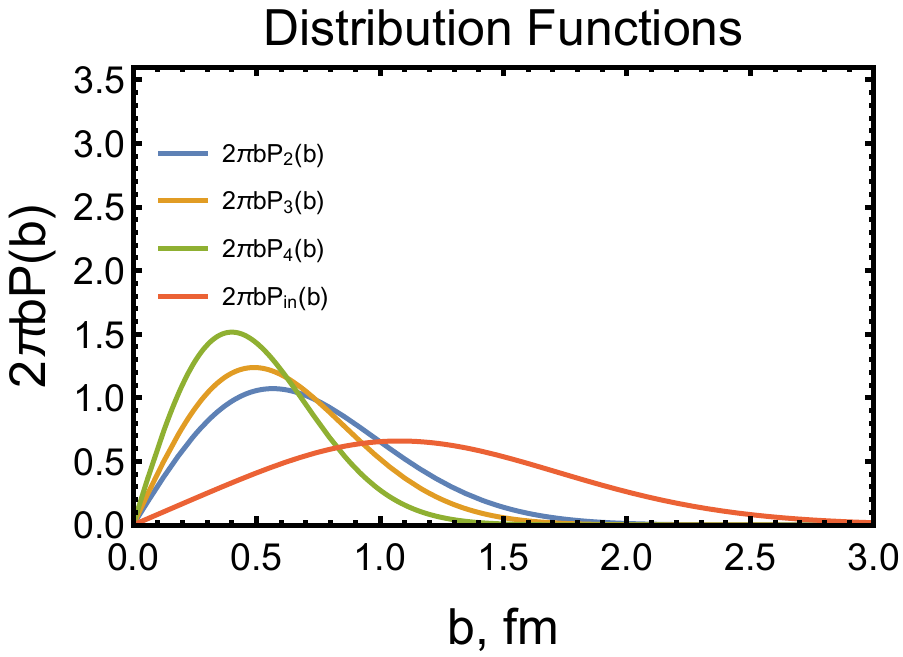}
\caption[]{Normalized probabilities of minimum bias, inclusive two, and four  parton  collisions and collision involving three partons as a function of the impact parameter.}
\label{scales}
\end{figure}

Our numerical studies indicate that  the
impact parameter distributions with the jet trigger (Eq.\ref{P_2_exp_dip})  are much
more narrow than that in minimum bias inelastic events at the same energy (Eq.\ref{P_in_def}) -- see Fig. ~\ref{scales},
and that  $b$-distribution for events with a dijet trigger is a very weak function of the $p_T$ of the jets or their rapidities.  For example, for the case of the $pp$ collisions at $\sqrt{s}= \mbox{13~GeV} $ the  median value of $b$,
$ b_{median} \approx$ 1.2 fm  and $ b_{median} \approx$
0.65 fm for minimum bias and dijet trigger events\cite{Frankfurt:2010ea}.

Note here that in many experimental analyses the minimum bias cross section is defined as the inelastic nondiffractive cross section.  Since inelastic diffraction is a peripheral process in $pp$ scattering, $\sigma_{min.bias} $ defined this way corresponds to   somewhat smaller $b$ than the ones given by Eq.~\ref{P_in_def}.

For $N\ge 2$ dijet processes
\be
b_{median}(N) \approx {1\over \sqrt{N}} b_{median}(N=1).
\label{central}
\ee
Hence inclusive $N \ge  2$ processes are dominated by collisions at very small impact parameters where  gluon fields of two nucleons  strongly overlap: $b_{median} < 2 r_g^{(N)} (x)$ (here   $r_g^{(N)} (x) \geq 0.4 fm$ is the transverse   radius of the gluon distribution in nucleons), cf. Fig.~\ref{scales}.

\par
Since the large impact parameters give the dominant contribution to $\sigma_{inel} $ our analysis indicates that
there are two pretty distinctive classes of $pp$ collisions  - large $b $ collisions which are predominantly soft and and central collisions with strongly enhanced rate of hard collisions. We refer to this pattern as the two transverse scale picture of $pp$ collisions at collider energies  \cite{FSW}.

 \section{GPD and mean field approach to MPI.}
Descryption of the MPI is a multi-scale problem. This  is not  only because the separate parton--parton interactions may differ in hardness.
More importantly, {\em each}\/ single hard interaction possesses {\em two}\/ very different hardness scales.
The distinctive feature of the DPS is that it produces two pairs of nearly back-to-back jets, so that
 in the collision of partons 1 and 3 the first (larger) scale is given by the invariant mass of the jet pair,
$Q^2 = 4J_{1\perp}^2 \simeq 4J_{3\perp}^2$, while
the second scale is the magnitude of the {\em total transverse momentum}\/ of the pair:
$\delta^2 = \delta_{13}^2$.
It is important to stress that in the MPI physics there is  {\em no factorization}\/ in the usual sense of the word.
The cross sections do not factorize into the product of the hard parton interaction cross sections and the multi-parton distributions depending on momentum fractions $x_i$ and the hard scale(s).
A general approach to double (multi) hard interactions has been developed in \cite{BDFS1}.
It turned out that the {\em transverse momentum}\/ of the parton in the w.f.\ and that of its counterpart in the conjugated w.f.\ are indeed necessarily different, with their difference $\vec{\Delta}$ being conjugate to the relative transverse distance between the two partons in the hadron.
This has led to introduction of  the new object -- {\em generalized double parton distribution}\/, $_2\mbox{GPD}$,
which depends on a new momentum parameter $\vec{\Delta}$ \cite{BDFS1,BDFS2}.

\subsection{Generalized two-parton distribution  \label{SEC:GPD}}
\subsubsection{$_2\mbox{GPD}$ and their connection to wave functions.}
In \cite{BDFS1,BDFS2} we have shown that the
QFT description of the double hard parton collisions calls for introduction of  $_2\mbox{GPD}$. Defined in the momentum space, it characterizes two-parton correlations inside hadron~\cite{BDFS1}:
$ D _h(x_1,x_2,Q_1^2,Q_2^2;\vec\Delta)$.
Here the index $h$ refers to the hadron, $x_1$ and $x_2$ are the light-cone fractions of the parton momenta, and $Q_1^2,Q_2^2$ the corresponding hard scales.
As has been mention above, the two-dimensional vector $\vec\Delta$ is the Fourier conjugate to the relative distance between the partons $1$ and $2$ in the impact parameter plane.
The distribution obviously depends on the parton species; we suppress the corresponding indices for brevity.

The $_2\mbox{GPD}$ are expressed through multiparton light cone wave functions as:
\begin{eqnarray}
&&\hspace{-1cm} D(x_1,x_2,p^2_1,p^2_2,\overrightarrow{\Delta})
=\sum_{n=3}^{\infty}\int
\frac{d^2k_1}{(2\pi)^2}\frac{d^2k_2}{(2\pi)^2}\theta
(p_1^2-k_1^2)\theta (p_2^2-k_2^2)
\nonumber\\[10pt] &&
\times \int \prod_{i\ne
1,2}\frac{d^2k_i}{(2\pi)^2}\int^1_0\prod_{i\ne
1,2} dx_i\, (2\pi)^3\delta( \sum_{i=1}^{i=n} x_i-1)
\delta (\sum_{i=1}^{i=n} \vec k_i)
\nonumber\\[10pt]
&&\hspace{-1cm}
\times
\psi_n (x_1,\vec k_1,x_2,\vec k_2,.,\vec k_i,x_i..)\psi_n^+(x_1,\overrightarrow{k_1}+\overrightarrow{\Delta},x_2,\overrightarrow{k_2}
-\overrightarrow{\Delta},x_3, \vec k_3,...)
.
\label{b2}
\end{eqnarray}
Note that this distribution is diagonal in the space of all
partons except the two partons involved
 in the collision. Here $\psi$ is the parton
wave function normalized to one in the  usual way. An appropriate
summation over color and Lorentz indices is implied.

The double hard interaction cross section (and, in particular, that of production of two dijets)
can be expressed through the  convolution of
$_2\mbox{GPD}$s.

 The
{\em effective interaction area}\/ $\sigma_\eff$ defined in Eq. \ref{sigmaeff}
is given by the convolution of  the $_2\mbox{GPD}$s of incident hadrons
over the transverse momentum parameter $\vec{\Delta}$ normalized by the product of single-parton inclusive pdfs:
\beq
\frac1{\sigma_\eff} \equiv \frac{
\int \frac{d^2\vec{\Delta}}{(2\pi)^2}  \> D_{h_1}(x_1,x_2, Q_1^2,Q_2^2;\vec\Delta) D_{h_2}(x_3,x_4, Q_1^2,Q_2^2; -\vec\Delta)}
{D_{h_1}(x_1,Q_1^2)D_{h_1}(x_2,Q_2^2)D_{h_2}(x_3,Q_1^2) D_{h_2}(x_4,Q_2^2)}.
\label{doublejet}
\eeq

Eq. 
\ref{doublejet} (and similar expression for any number of MPI)
can be rewritten in transverse coordinate  representation and corresponds to the  transverse geometry depicted in
 Fig.~\ref{overlap4} with $\vec{\Delta} $ Fourier conjugated to the difference of transverse coordinates  of partons: $\vec{\rho}_1-\vec{\rho}_3$.

\par
$_2\mbox{GPD}$s enter also the expressions for the differential distributions in the jet transverse momentum imbalances $\vec{\delta}_{ik}$
(integral of which  over $\vec{\delta}_{ik}$ is  the ``total'' DPS cross section -- Eq.\ref{doublejet}.
In the inclusive case the hardness parameters of the $_2\mbox{GPD}$s are given by the jet transverse momenta $Q_i^2$,
while for  the differential distributions --- by the jet imbalances $\delta_{ik}^2$.
The corresponding formulae  derived in the leading collinear approximation of pQCD can be found in Ref.~\cite{BDFS2}.
 It is worth emphasizing here  that the DPS cross section {\em does not factorize}\/ into the product of the hard parton interaction cross sections and the two two-parton distributions depending on momentum fractions $x_i$ and the hard scales, $Q_1^2, Q_2^2$.

  Note that  one can introduce in the same way the $N$-particle GPD, $G_N$,
which can be probed in the production of $N$ pairs of jets \cite{BDFS1}.
 In this case  the   first $N$ arguments $k_i$
  are shifted by
$\overrightarrow{\Delta_i}$ subject to the  constraint $\sum_i \overrightarrow{\Delta_i}=0$. So the cross
section is proportional to
\begin{eqnarray}
\sigma_{2N}&\propto& \int \prod_{i=1}^{i=N}{d\overrightarrow{\Delta}_i\over (2\pi)^2}
D_a(x_1,... x_N, \overrightarrow{\Delta}_1,...\overrightarrow{\Delta}_N)
\nonumber\\[10pt]
 &\times&
 D_b(x_1',... x_N',\overrightarrow{\Delta}_1,...\overrightarrow{\Delta}_N)
\delta(\sum_{i=1}^{i=N}\overrightarrow{\Delta}_i).
\label{multi}
\end{eqnarray}
N-parton GPD are expressed through multiparton wave functions  analogously  to Eq.\ref{b2}.

\par The above approach
 allows to take into account consistently the perturbative mechanism of two-parton correlation
 when the two partons emerge from {\em perturbative splitting}\/ of one parton taken from the hadron wave function since one needs to separate these correlations from the $2\to 4$ mechanism of jet production.

In perturbative scenario the production of the parton pairs is concentrated at much smaller transverse distances between partons.
As a result, the corresponding contribution to $_2\mbox{GPD}$.
 turns out to be practically independent of $\Delta^2$
in a broad range, up to the hard scale(s) characterizing the hard process under consideration
($\Delta^2$ only affects the lower limit of the transverse
momentum integrals in the parton cascades,
resulting in
a mild logarithmic dependence).
The weak dependence on $\Delta$ results in a distribution over impact parameters for DPS which is intermediate between the mean field contribution  and dijet b - distributions, cf. Fig.~\ref{scales}.
\begin{figure}[htbp]
\hspace{1.7cm}\includegraphics[scale=0.5]
{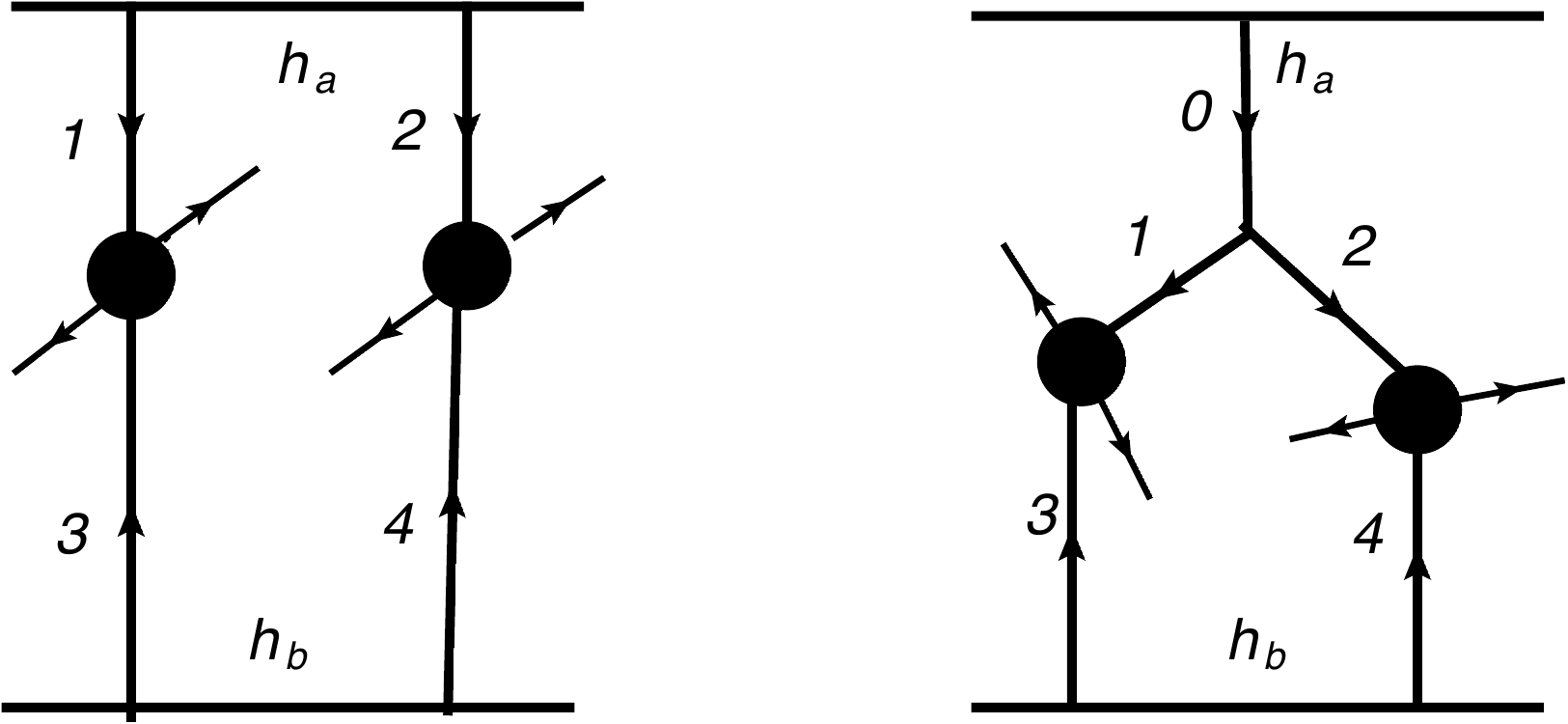}
\caption{Sketch of the two considered DPS mechanisms: \22 (left) and \12 (right) mechanism.}
\label{fig1a}
\end{figure}
Given essentially different dependence on $\Delta$, one has to treat the two contributions separately by casting the $_2$GPD as a sum of two terms  depicted in Fig. ~\ref{fig1a}:
\beq
D_h  (x_1,x_2, Q_1^2,Q_2^2;\vec\Delta) =
{}_{[2]}D_h(x_1,x_2, Q_1^2,Q_2^2;\vec\Delta)
 + {}_{[1]}D_h(x_1,x_2, Q_1^2,Q_2^2;\vec\Delta) .
 \label{eq:2terms}
\eeq
Here subscripts ${}_{[2]}D$ and ${}_{[1]}D$ mark the first and the second mechanisms, correspondingly:
two partons from the wave function versus one parton that perturbatively splits into two (see Fig.~\ref{fig1a})
\par  \par Let us stress that it
follows from the above formulas that in the impact parameter space  these GPDs have a
 probabilistic interpretation. In particular  they are positively definite  in the impact parameter space, see discussion in \cite{BDFS4}.

\subsection{Modeling ${}_{[2]}\mbox{D}$: the mean field approach.\label{SEC:Model}}
To proceed with quantitative estimates, one needs a model for the non-perturbative two-parton distributions in a proton.
A priori, we know next to nothing about them.
The first natural step to take is an {\em approximation of independent partons}/mean field approximation.
It allows one to relate $_2$GPD with known objects, namely~\cite{BDFS1}

\beq \label{eq:DGG}
{}_{[2]}D(x_1,x_2, Q_1^2,Q_2^2;\Delta) \simeq G(x_1,Q_1^2;\Delta^2) G(x_2, Q_2^2;\Delta^2).
\eeq
Here $G$ is the non-forward parton correlator (known as generalized parton distribution, GPD)
that determines, e.g., hard vector meson production at HERA
and which enter in our case in the diagonal kinematics in $x$ ($x_1=x_1'$).

Modeling
$_2$GPD  using Eq.~\ref{eq:DGG} has its limitations. First of all, it does not respect the  obvious restriction $D(x_1+x_2> 1)=0$. So, $x_i$ have to be taken not too large (say, $x_i\ll 0.5$). Actually, the neglect of correlations is likely to be a good approximation only at much smaller $x\le 0.1$.
In any case, currently one can extract GPDs only from the theoretical analysis of the hard exclusive amplitude like $\gamma^*_L +  N\to VM +N$ are only available for $x < 0.05$.

There is an additional caveat - in the vector meson production
 two gluons in t-channel carry different light cone fractions while in the case of the scatering amplitude x's in the $\left|in\right>$ and $\left<out\right|$ state are equal.
 Also, in the vector meson production modulus squared of the amplitude enters while in our case we deal with the imaginary part of the zero angle amplitude. As a result a simple connection between the gluon GPD and the observed cross section exists only if virtualities are large enough and $x$ is small enough, $x\le 0.1$.

On the other hand, $x_i$ should not be {\em too small}\/ to stay away from the region of the Regge-Gribov phenomena where there are serious reasons for parton correlations to be present at the non-perturbative level (see discussion in \cite{BDFS3} and in section 6).

Thus, we expect that $x$-range where the mean field NP
model \ref{eq:DGG} is applicable for $_2\mbox{GPD}$
is $10^{-1}\geq x_i \geq 10^{-3}$.

The GPDs can be parameterized  as
\beq \label{eq:GDF}
 G(x_1,Q_1^2;\Delta^2) \>\simeq\> D(x_1, Q_1^2) \times F_{2g}(x_1,\Delta^2, Q^2) ,
\eeq
with $D$  being the usual one-parton distribution functions
and $F$ being the so-called two-gluon form factor of the hadron.
The latter is a non-perturbative object; it falls fast with the ``momentum transfer'' $\Delta^2$.
In our following numerical studies we will use the model of the two gluon form factor extracted from   the  data on exclusive $J/\psi$ photoproduction. This analysis is summarized in the  Appendix.

Using parametrization of
Eq.~(\ref{P_2_exp_dip}) one finds
\cite{FSW,BDFS1}
\beq \frac{1}{\sigma_{eff}}=\int
\frac{d^2\Delta}{(2\pi)^2}F_{g}^4(\Delta)(8\pi B_g)^{-1}\approx  32 mb,
\label{b4}
\end{equation}
for $x\sim 0.01$
for the exponential parametrization  fit and and practically the same number, $\frac{m^2_g}{28\pi}$, for dipole fit with  $B_g$ related to $m_g^2$ according to Eq.\ref{dip_exp}.
Numerically Eq.\ref{b4}
   leads to approximately a factor of two smaller production  cross section than the one observed at the Tevatron at $x\ge 0.01$.
   Since the two gluon form factor decreases faster with $t$  with decrease of $x$, the mean field model leads to increase of \effs with energy for the central rarities and fixed $p_t$.
 Note that the two exponential parametrization of transverse parton density used in a number of versions of Pythia which described experimental values of \effs strongly contradicts the data on the $J/\psi$ photoproduction, see e.g. Fig.~3  in \cite{Zakopane}.

Using Eq.\ref{multi} and exponential parametrization of GPD  one can also find the effective cross section  for $n$ hard collisions
in mean field approach:
 \beq
 {1\over \sigma_{eff}^{(n)}}= \frac{1}{(2\pi)^{N-1}}\prod_{i=1}^{i=N}\frac{1}{B_i+B_i'}\frac{1}{\sum_{i=1}^{i=N}1/(B_i+B_i')}.
 \eeq
Here $B_i\equiv B(x_i), B'_i\equiv B(x_i')$ for N dijet process with $x_i$ are Bjorken fractions for hadron a, and $x_i'$ are Bjorken fractions for colliding hadron   b. For N=2 we get the  familiar result \cite{BG1}:
\beq
\frac{1}{\sigma_{\rm eff}}=\frac{1}{2\pi}\frac{1}{B_1+B_1'+B_2+B_2'}.
\eeq
The particular case of this formula for N=3 was  recently
 considered
 in \cite{Enterria}.

\section{pQCD correlations.}

\subsection{ \12\ DPS process \label{SEC:12}}

Actually, the NP and PT contributions {\em do not}\/ enter the physical DPS cross section in the arithmetic sum Eq.\ref{eq:2terms}, driving one even farther from the familiar factorization picture based on universal (process independent) parton distributions.
As explained in \cite{BDFS2},  a double hard interaction of two pairs of partons that {\em both}\/ originate from PT
splitting of a single parton from each of the colliding hadrons, does not produce back-to-back dijets.
In fact, such an eventuality corresponds to a one-loop correction to the usual $2\to4$ jet production process
and should not be looked upon as a multi-parton interaction.
The term ${}_{[1]}D_{h_1}\times {}_{[1]}D_{h_2}$ has to be excluded from the product $D_{h_1}\times D_{h_2}$,
the conclusion we share with Gaunt and Stirling \cite{stirling1}.

So, we are left with two sources of genuine two-parton interactions:
four-parton collisions described by the product of (PT-evolved) $_2$GPDs of NP origin (\22),

\beq\label{eq:22}
{}_{[2]}\!D_{h_1}(x_1,x_2, Q_1^2,Q_2^2;\vec\Delta)\,  { } _{[2]}\!D_{h_2}(x_3,x_4,  Q_1^2,Q_2^2; -\vec\Delta) ,
\eeq
and three-parton collisions due to an interplay between the NP two-parton correlation in one hadron
and the two partons emerging from a PT parton splitting in another hadron (\12\ ), described by the combination
\begin{eqnarray}
& &  _{[2]}\!D_{h_1} (x_1,x_2, Q_1^2,Q_2^2; \vec\Delta)
\, _{[1]}\!D_{h_2} (x_3,x_4, Q_1^2,Q_2^2; -\vec\Delta)\nonumber
\\[10pt]
  \hspace{-1cm}    &+ &\>        {}_{[1]}\!D_{h_1} (x_1,x_2, Q_1^2,Q_2^2; \vec\Delta)
\, _{[2]}\!D_{h_2}(x_3,x_4, Q_1^2,Q_2^2;-\vec\Delta) .
\label{eq:12}
\end{eqnarray}

Given that ${ }_{[2]}\!D$ falls fast at large $\Delta$, a mild logarithmic $\Delta$-dependence of ${}_{[1]}D$ can be neglected in the product in Eq.~\ref{eq:12}.

\subsection{Composition of the \12 DPS cross section}
In order to derive the DPS cross section, one has to start with examination of the double differential  transverse momentum distribution and then integrate it over jet imbalances $\delta_{ik}$.
Why this step is necessary?
The parton distribution  $D(x,Q^2)$ --- the core object of the QCD-modified parton model ---  arises upon logarithmic integration over the transverse momentum up to the hard scale, $k_\perp^2< Q^2$.
Analogously, the double parton distribution $D(x_1,x_2,Q_1^2,Q_2^2; \vec\Delta)$ embeds {\em independent integrations}\/ over parton transverse momenta $k_{1\perp}^2$, $k_{2\perp}^2$ up to $Q_1^2$ and $Q_2^2$, respectively.
However, the \12\ DPS cross section contains a specific contribution ("short split'', see below) in which the transverse momenta of the partons 1 and 2 are strongly correlated (nearly opposite). This pattern does not fit into the structure of the pQCD evolution equation for ${ }_2\mbox{GPD}$ where $k_{1\perp}$ and $k_{2\perp}$ change independently.
Given this subtlety, a legitimate question arises whether the expression for the integrated \12\ cross section Eq.\ref{eq:12} based on the notion of the two-parton distribution ${ }_{[1]}\!D$
takes the short split into account.
The differential distribution over jet imbalances was derived in \cite{BDFS2} in the leading collinear approximation of pQCD.
It resembles the ``DDT formula'' for the Drell-Yan spectrum \cite{DDT} and
contains two derivatives of the product of $_2\mbox{GPD}$s Eq.\ref{22}
that depend on the corresponding $\delta_{ik}$ as hardness scales, and the proper Sudakov form factors depending on (the ratio of) the $Q_i^2$ and $\delta^2_{ik}$.

In particular,  in the region of {\em strongly ordered}\/ imbalances,
\beq\label{eq:dlog1}
\frac{\pi^2 d\sigma^{\mbox{\scriptsize DPS}}}{d^2\delta_{13} \, d^2\delta_{24}} \propto \frac{\as^2}{\delta_{13}^2 \, \delta_{24}^2}; \quad
\delta_{13}^2 \>\gg\> \delta_{24}^2, \>\>   \delta_{13}^2 \>\ll\> \delta_{24}^2,
\eeq
the differential \12\ cross section reads
\begin{eqnarray}
\frac{\pi^2 d\sigma_{\mbox{\scriptsize\12}}}{d^2\delta_{13}\, d^2\delta_{24}}& =&
 \frac{d\sigma_{{\mbox{\scriptsize part}}}}{d\hat{t}_1\,d\hat{t}_2}
 \frac{d}{d\delta_{13}^2} \frac{d}{d\delta_{24}^2} \bigg\{   \int\!\! \frac{d^2\vec{\Delta}}{(2\pi)^2} \>\nonumber\\[10pt]
&\times&{}_{[1]}\!D_{h_1}(x_1,x_2,\delta_{13}^2,
 \delta_{24}^2; \vec{\Delta})
 \> {}_{[2]}\!D_{h_2}(x_3,x_4,\delta_{13}^2, \delta_{24}^2; \vec{\Delta} )\nonumber\\[10pt]
&\times& S_1\left({Q_1^2},\delta_{13}^2 \right)S_3\left({Q_1^2},{\delta_{13}^2}\right)
 \cdot S_2\left({Q_2^2},{\delta_{24}^2}\right) S_4\left({Q_2^2},\delta_{24}^2\right)   \bigg\}\nonumber\\[10pt]
  &+ & \big\{ h_1 \leftrightarrow h_2 \big\}.
 \label{22}
\end{eqnarray}
The differential distribution for the \22\ DPS mechanism has a similar structure, see Eq.~(25) of \cite{BDFS2}.

In addition to Eqs.~\ref{eq:dlog1},\ref{22} there is another source of double collinear enhancement in the differential \12\ cross section.
It is due to the kinematical region where the two imbalances nearly compensate each other,
\beq\label{eq:deltaprime}
\delta'^2 = (\vec{\delta}_{13} + \vec{\delta}_{24})^2 \>\ll\> \delta^2 = {\delta}_{13}^2\simeq {\delta}_{24}^2,
\eeq
and the dominant integration region is complementary to that of Eq.\ref{eq:dlog1}:
\beq\label{eq:dlog2}
\frac{\pi^2 d\sigma^{\mbox{\scriptsize DPS}}_{\mbox{\scriptsize short}}}{d^2\delta_{13} \, d^2\delta_{24}} \propto \frac{\as^2}{\delta'^2 \, \delta^2}; \qquad
\delta'^2 \>\ll\> \delta^2 .
\eeq
This enhancement characterizes the set of \12\  graphs in which accompanying radiation
has transverse momenta not exceeding $\vec\delta'$.

\noindent
In this situation, the parton that compensates the overall imbalance,
$\vec{k}_\perp= - \vec{\delta}'$ is radiated off the incoming, quasi-real, parton legs.
At the same time, the virtual partons after the core splitting ``0''$\to$ ``1''+``2'' enter their respective hard collisions without radiating any offsprings on the way.

The  $1\to2$ splitting 
occurs close to
 the hard vertices, therefore the name "short split'' (aka "endpoint contribution", \cite{BDFS2}).

A complete
expression for the differential distribution in  the jet imbalances due to a short split was  derived in the leading collinear approximation (Eq.~(27) of \cite{BDFS2}):
\begin{eqnarray}
\label{eq:DDT32}
 \frac{\pi^2\> d\sigma^{\mbox{\scriptsize DPS}}_{\mbox{\scriptsize short}}}{d^2\delta_{13}\, d^2\delta_{24}} \>&=&\>\>
 \frac {d\sigma_{{\mbox{\scriptsize part}}}
 } {d\hat{t}_1\,d\hat{t}_2}\> \cdot \>   \frac{\as(\delta^2)}{2\pi\, \delta^2} \,  \sum_c  P_{c}^{(1,2)}\!\!
    \left(\frac{x_1}{x_1+x_2}\right) \nonumber\\[10pt]
&\times&    S_1(Q_1^2,\delta^2)\, S_2(Q_2^2,\delta^2)\nonumber\\[10pt]
{ }\quad &\times& \frac{d}{d\delta'^2}
 \bigg\{ S_c(\delta^2\!,\delta'^2) \frac{D_{h_1}^{c}(x_1\!+\! x_2,\delta'^2)}{x_1+x_2}
 S_3(Q_1^2 ,\delta'^2) S_4(Q_2^2,\delta'^2) \!\nonumber\\[10pt]
&\times&  \int\!\! \frac{d^2\vec{\Delta}}{(2\pi)^2} \, {}_{[2]}\!D_{h_2}(x_3,x_4,\delta'^2\! , \delta'^2;\vec{\Delta} ) \bigg\}
+ \big\{ h_1 \leftrightarrow h_2 \big\}.
\end{eqnarray}

The short split becomes less important when the scales of the two hard collisions
are different.
Indeed, the logarithmic integration over $\delta^2$ is kinematically restricted from above,
$\delta^2 < \delta^2_{\mbox{\scriptsize{max}}} \simeq \min\{ Q_1^2, Q_2^2\}$.
As a result, in the kinematics where  transverse momenta of jets in one pair are much larger than in
the second pair, e.g.,
$ Q_1^2 \gg Q_2^2$ ,
the contribution of the short split is
suppressed as
\[
  \left.
  {\sigma^{(3\to4)}_{\mbox{\scriptsize short}}}  \right/  {\sigma^{(3\to4)}} \>\propto\> S_1(q_1^2,q_2^2)\, S_3(q_1^2,q_2^2)  \> \ll\> 1
  \quad (Q_1^2\gg Q_2^2).
\]
Here $S_1$ and $S_3$ are the double logarithmic Sudakov form factors of the partons ``1'' and ``3'' that enter the hard interaction with the larger hardness scales.
The short split induces a strong correlation between jet imbalances which is worth trying to look for experimentally.

The relative weight of the short split depends on the process under consideration.
For most DPS processes in the kinematical region we have studied, it typically provides 10--15\%\ of
the pQCD correlation contribution.
However, it
becomes more important when the nature of the process favors parton splitting. In particular,  this is the case for
 the double Drell-Yan pair production where the short split contribution reaches 30--35\%.
On the contrary, the short split  turns out to be practically negligible for the same-sign double $W$-meson production \cite{BDFS4}.

Thus, for the integrated DPS cross section we obtain two contributions to the effective interaction area:
\begin{eqnarray}
\label{eq:eff4}
\frac{\prod_{i=1}^4 \! D(x_i)}{\sigma_{4}} \!\!\!&=&\!\!\!\!
\int\! \frac{d^2\vec{\Delta}}{(2\pi)^2} \>
 _{[2]}D_{h_1}(x_1,x_2, Q_1^2,Q_2^2;\vec\Delta)
\> {}_{[2]}D_{h_2}
(x_3,x_4, Q_1^2,Q_2^2; -\vec\Delta),\nonumber\\[10pt]
\frac{\prod_{i=1}^4 \! D(x_i)}{\sigma_{3}} \!\!\!&=&\!\!\!\!\!
\int\!\! \frac{d^2\vec{\Delta}}{(2\pi)^2}  \! \bigg[ {}_{[2]}
\!D_{h_1}\!
(x_1,x_2, Q_1^2,Q_2^2; \!\vec\Delta) {}_{[1]}\!D_{h_2}\!
(x_3,x_4, Q_1^2,Q_2^2)\nonumber\\[10pt]
&+&\! {} _{[1]}\!D_{h_1}\!
(x_1,x_2, Q_1^2,Q_2^2) {}_{[2]}\!D_{h_2}\!
(x_3,x_4, Q_1^2,Q_2^2;  \!\vec\Delta)\!\bigg]\!. \qquad {  }
\end{eqnarray}

Let us stress here that our analysis demonstrates  that a compact and intuitively clear expression containing the product of the $_2$GPDs ${}_{[2]}\mbox{D}$ and ${}_{[1]}\mbox{D}$ in Eq.\ref{eq:eff4} is valid
only for  the {\em integrated}\/ \12\ cross section.

\subsection{Modeling $_1D$ terms.}

Turning to the \12\ term, we neglect a mild logarithmic $\Delta$-dependence of $_{[1]}D$ in \ref{eq:eff4} and use the model of section 3B for $_{[2]}D$  to obtain
\beq
 {\sigma_{3}}^{-1}  \>\simeq\>
 \frac73\cdot \left[ \frac{_{[1]}D(x_1,x_2)}{D(x_1)D(x_2)} + \frac{_{[1]}D(x_3,x_4)}{D(x_3)D(x_4)} \right] \times {\sigma_{4}}^{-1} ,
\eeq
where we substituted the value of the integral
\[
     \int\! \frac{d^2\vec{\Delta}}{(2\pi)^2} \> F_{2g}^2(\Delta^2) \>=\>\frac{m_g^2}{12\pi}.
\]
\par Very similar results are obtained for expomential parametrisation.

We will parametrize the result in terms of the ratio
 \beq
      R\>\equiv\>  \frac{\sigma_{\mbox{\scriptsize \12}}}{\sigma_{\mbox{\scriptsize \22}}} \>=\> \frac{\sigma_4}{\sigma_3}.
      \label{eq:Rdef}
 \eeq
For the effective interaction area,
\beq\label{eq:Sfull}
 \sigma_\eff^{-1} \>=\> \sigma_{4}^{-1} + \sigma_{3}^{-1},
\eeq
we parametrize
\beq
\sigma_{eff}=\frac{\sigma_{\rm eff}^{\rm mean\,\,\, field}}{1+R},
\label{Rmean}
\eeq
where $\sigma_{eff}^{mean field}$ is the mean field value of \effs, obtained either using dipole or exponential fit.
The difference between the values for \effs obtained using these two fits is within current   experimental errors of the $J/\psi $ data.
In numerical simulations for DPS below we use dipole fit, that works slightly better for values of Bjorken x corresponding to hard DPS, while for the  underlying event (UE) we used the exponential fit, that works slightly better for small x relevant for the UE.
The difference  however is of the order of  several percent and can be neglected.

Within the framework of the NP
two-parton $_2\mbox{GPD}$ model,  Eq.\ref{eq:DGG},
there is
only one free parameter $Q_0^2$.
The DPS theory can be applied to various processes and holds in a range of energies and different kinematical regions.
Therefore, having fixed the $Q_0^2$ value, say, from the Tevatron data, one can consider all other applications (in particular, to the LHC processes) as parameter-free theoretical predictions.

\subsection{Analytical estimate of pQCD correlations.}
 The PT parton correlations cannot be
 neglected.
 Indeed, let us chose a scale $Q_{0}$ that separates NP and PT physics to be sufficiently low, so that parton cascades due to the evolution between $Q_{0}$  and $Q_i^2$  are  well developed.
To get a feeling of the relative importance of the PT correlation, as well as to understand its dependence on $x$ and the ratio of scales, $Q^2$ vs \ $Q_{0}^2$, the following lowest order PT estimate can be used.

Imagine that at the scale $Q_{0}$ the nucleon consisted of $n_q$ quarks and $n_g$ gluons ("valence partons'') with relatively large longitudinal momenta,
so that triggered partons with $x_1,x_2\ll1$ resulted necessarily from PT evolution.
In the first logarithmic order,
  $\as\log (Q^2/Q_{0}^2)\equiv \xi$,
 the inclusive spectrum can be represented as
\[
    D \propto (n_qC_F + n_gN_c)\xi,
\]
where we suppressed $x$-dependence as irrelevant.
If both gluons originate from the same ``valence'' parton, then

\beq \label{eq:1D}
{}_{[1]}\!D \propto  \frac12N_c\xi\cdot D \>+\,  (n_qC_F^2 + n_gN_c^2)\xi^2,
\eeq
while independent sources give: 

\beq \label{eq:2D}
{}_{[2]}\!D \propto \big(n_q(n_q\!-\!\!1)C_F^2 + 2n_qn_gC_FN_c +n_g(n_g\!-\!\!1)N_c^2\big)\xi^2
 = D^2 - \big(n_qC_F^2 + n_gN_c^2\big)\xi^2 .
\eeq
Hence
\beq
c{D(x_1,x_2;0)}{D(x_1) D(x_2)} -1 \>\simeq\> \frac{N_c}{2(n_qC_F+n_gN_c)}.
\eeq
The correlation is driven by the gluon cascade ---- the first term in Eq.\ref{eq:1D} --- and is not small (being of the order of unity).
It gets diluted when the number of independent ``valence sources'' at the scale $Q_{02}$ increases.
This happens, obviously, when $x_i$ are taken smaller. On the other hand, for large $x_i\sim 0.1$ and increasing, the effective number of more energetic partons in the nucleon is about two and decreasing, so that the relative importance of the \12 processes grows.

We conclude that the relative size of PT correlations is of the order one, provided $\xi=\cO{1}$.
\medskip
  \section{Numerical results for DPS \label{SEC:Numerics}}

\subsection{Calculation framework}
We consider in this chapter hard DPS with $p_t>10 \div 15$ GeV, where the
\12 mechanism gives the dominant correction to the  mean field results.(The pattern for small $p_t$ relevant for UE is discussed in the  next section).
In numerical calculations we used the GRV92 parametrization of gluon and quark parton distributions in the proton \cite{GRV}.
We have checked that using more advanced GRV98 and CTEQ6L parametrizations does not change the numerical results.
The explicit GRV92 parametrization is speed efficient and allows one to start the PT
evolution
at rather small virtuality scales.
The combination $(Q_0^2+\Delta^2)$ was used as the lower cutoff for the logarithmic transverse momentum integrals in the parton evolution,
which induced a mild (logarithmic) $\Delta$-dependence on top of the relevant power of the two-gluon form factor $F_{2g}(\Delta^2)$.

To quantify the role of the \12\  DPS subprocesses, we calculated the ratio $R$ defined in  Eq.~\ref{eq:Rdef}
in the kinematical region $10^{-3} \leq x_i \leq 10^{-1}$ for Tevatron ($\sqrt{s}=1.8 \div 1.96 \, \mbox{TeV}$) and LHC energies ($\sqrt{s}= 7\, \mbox{TeV}$).
We chose to consider three types of  ensembles of colliding partons:
\begin{enumerate}
\item
 $u(\bar{u})$ quark and three gluons which is relevant for ``photon plus 3 jets'' CDF and D0 experiments,
\item
four gluons (two pairs of hadron jets),
\item
$u\bar{d}$ plus two gluons, illustrating  $W^+jj$ production.
\item
$u\bar{d}$ plus $d\bar{u}$, corresponding to the $W^+W^-$ channel.
\end{enumerate}
\medskip
 \subsection{Perturbative \12 correlation at the Tevatron.}
\subsubsection{CDF experiment}

In Fig.~\ref{FigCDF10} we show the profile of the  \12\ to \22\ ratio $R$ for the $\gamma+3$jets process
in the kinematical domain of the CDF experiment \cite{Tevatron1}.
The calculation was performed for the dominant ``Compton scattering'' channel of the photon production: $g(x_2)+u(\bar{u})(x_4)\to \gamma + u(\bar{u})$.
The longitudinal momentum fractions of two gluons producing second pair of jets are $x_1$ and $x_3$.
The typical transverse momenta were taken to be $p_{\perp 1,3}\simeq 5\,\GeV$ for the jet pair,
and $p_{\perp 2,4}\simeq 20\, \GeV$ for the photon--jet system.
In Fig.~\ref{FigCDF10} $R$ is displayed as a function of rapidities of the photon--jet,  $\eta_2 = \half\ln(x_2/x_4)$, and the 2-jet system,
$\eta_1 = \half\ln(x_1/x_3)$.
\begin{figurehere}
\begin{center}
\includegraphics[scale=0.6]{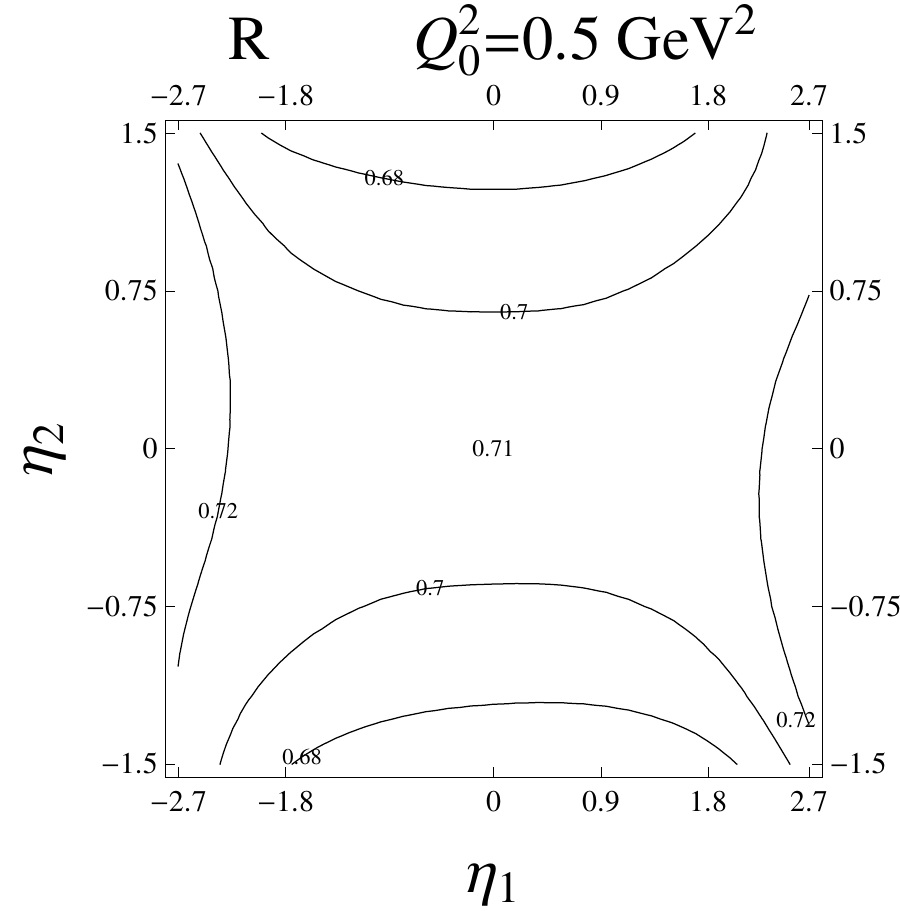}
\includegraphics[scale=0.6]{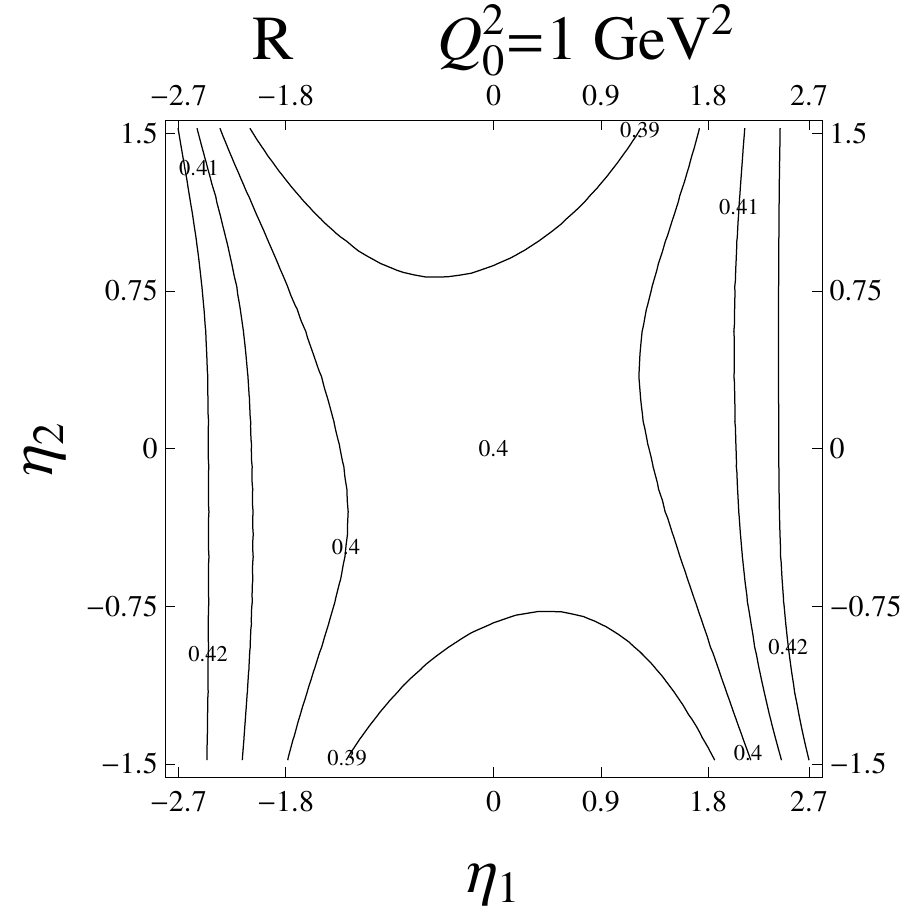}
\caption{\label{FigCDF10} The  \12/\22\ ratio, Eq.\ref{eq:Rdef} in the CDF kinematics
for the process $p\bar p \to \gamma + 3\, \mbox{jets}  + X$.}
\end{center}
\end{figurehere}

We observe that the enhancement factor lies in the ballpark of $1+R\sim 1.5\div 1.8$. Processed through
Eq.~\ref{Rmean},
it translates into $\sigma_\eff \simeq 18\div 21$~mb. This expectation has to be compared with the CDF finding
$\sigma_\eff =14.5 \pm 1.7\>^{+1.7}_{-2.3}\>\mbox{mb}$.
A recent reanalysis of the CDF data  points at an even small value: $\sigma_\eff =12.0\pm 1.4\>^{+1.3}_{-1.5}\>\mbox{mb}$,  \cite{Sjodmok}. Both these values are significantly smaller than our estimate and the result of D0 experiment discussed in the next subsection.

The results of numerical calculation for a fixed hardness $Q^2$ are shown in Fig.~\ref{FigCDF10}
for the CDF kinematics. We find that the $R$ factor and hence $\sigma_\eff$
exhibits a very mild $x$-dependence.

\subsubsection{D0 experiment}

The ratio $R$ is practically constant in the kinematical domain of the D0 experiment which studied
photon+3 jets production \cite{Tevatron2,Tevatron3} and is very similar to that of the CDF experiment shown above in Fig.~\ref{FigCDF10}.
So, for the D0 kinematics we instead display in Fig.~\ref{FigD0pt1}
the enhancement factor $1\!+\!R$ as a function of  $p_\perp$ of the secondary jet pair
for  photon transverse momenta 10, 20, 30, 50, 70, and 90 GeV (from bottom to top).
\begin{figurehere}
\begin{center}
\includegraphics[scale=0.45]{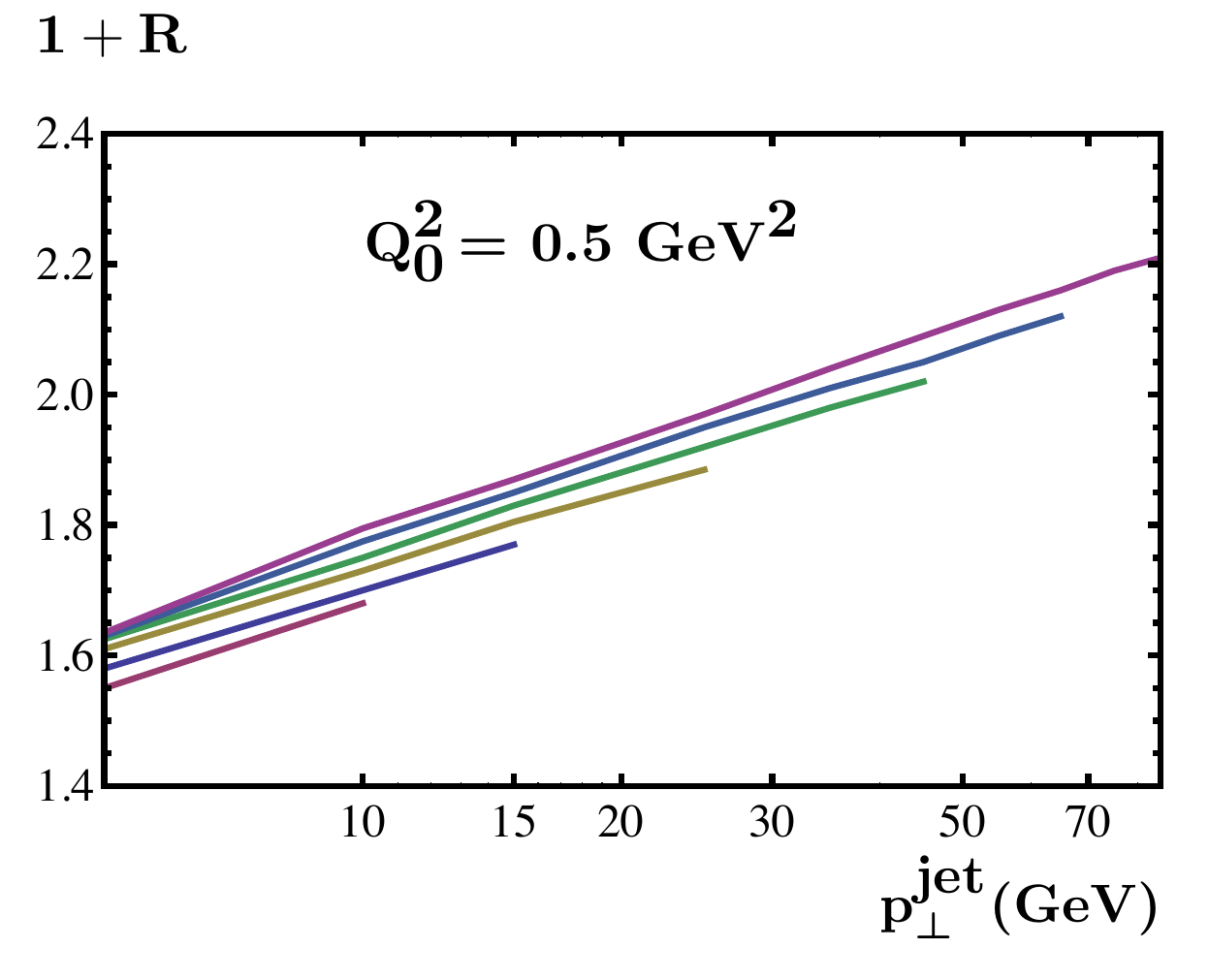}
\includegraphics[scale=0.45]{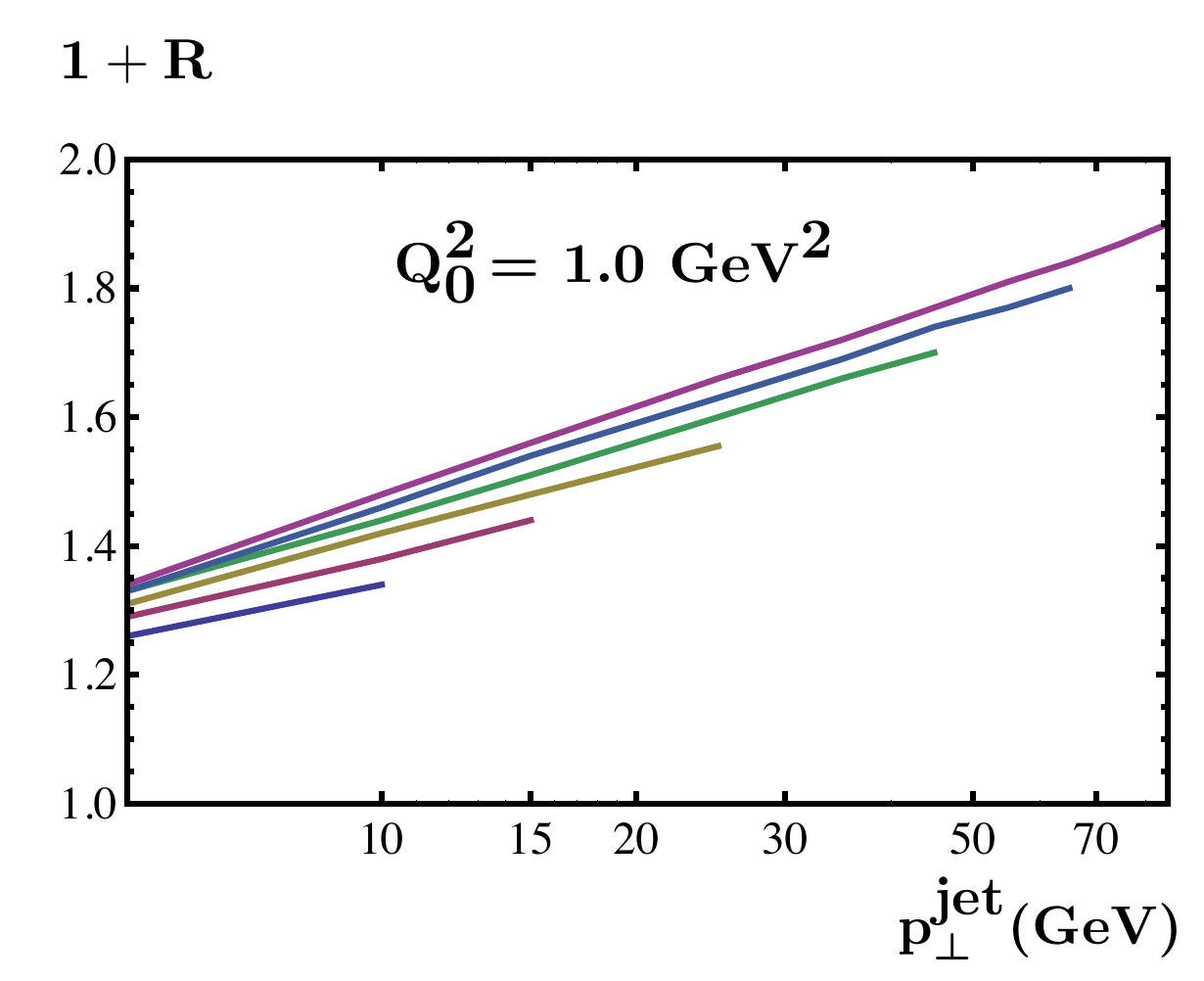}
\caption{\label{FigD0pt1}left: Central rapidity photon+3 jets production in $u$($\bar{u}$)--gluon collisions
in the D0 kinematics for $Q_0^2=0.5\GeV^2$. Right:  for $Q_0^2=1\GeV^2$. }
\end{center}
\end{figurehere}
The corresponding prediction for $\sigma_\eff$ is shown in Fig.~\ref{FigD0Seff} in comparison with the D0 findings.
\begin{figurehere}
\begin{center}
\includegraphics[scale=0.8]{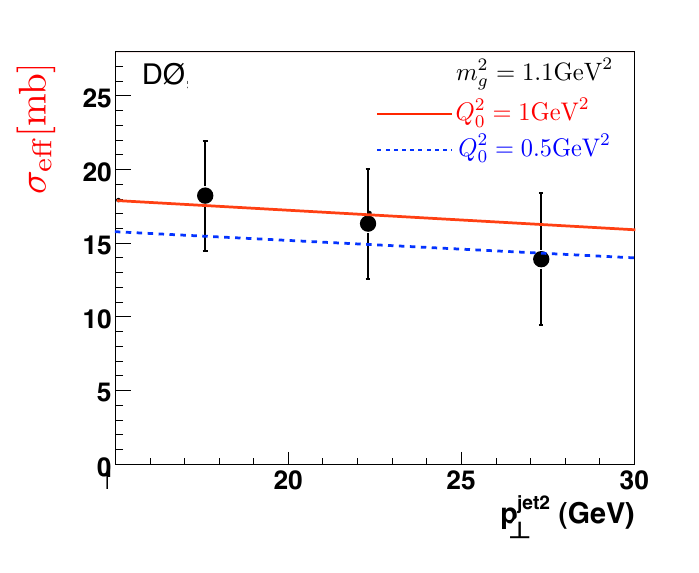}
\caption{ \label{FigD0Seff} $\sigma_\eff$ as a function of the hardness of the second jet in the kinematics of the D0 experiment \cite{Tevatron2,Tevatron3} for $p_{\perp\gamma}=70\,\GeV$.}
\label{D0plot}
\end{center}
\end{figurehere}
Both the absolute value and  the  hint of decrease of $\sigma_{eff}$ with increase of   $p_\perp$ look satisfactory.

\subsection{LHC energies}

In Fig.~\ref{FigLHCgg05} we show the \12\ to \22\ ratio for production of two pairs of back-to-back jets with transverse momenta $50\,\GeV$ produced in collision of gluons at the LHC energy of  $\sqrt{s}=$ 7\, TeV (the pattern for higher LHC energies is very similar).
\begin{figurehere}
\begin{center}
\includegraphics[scale=0.6]{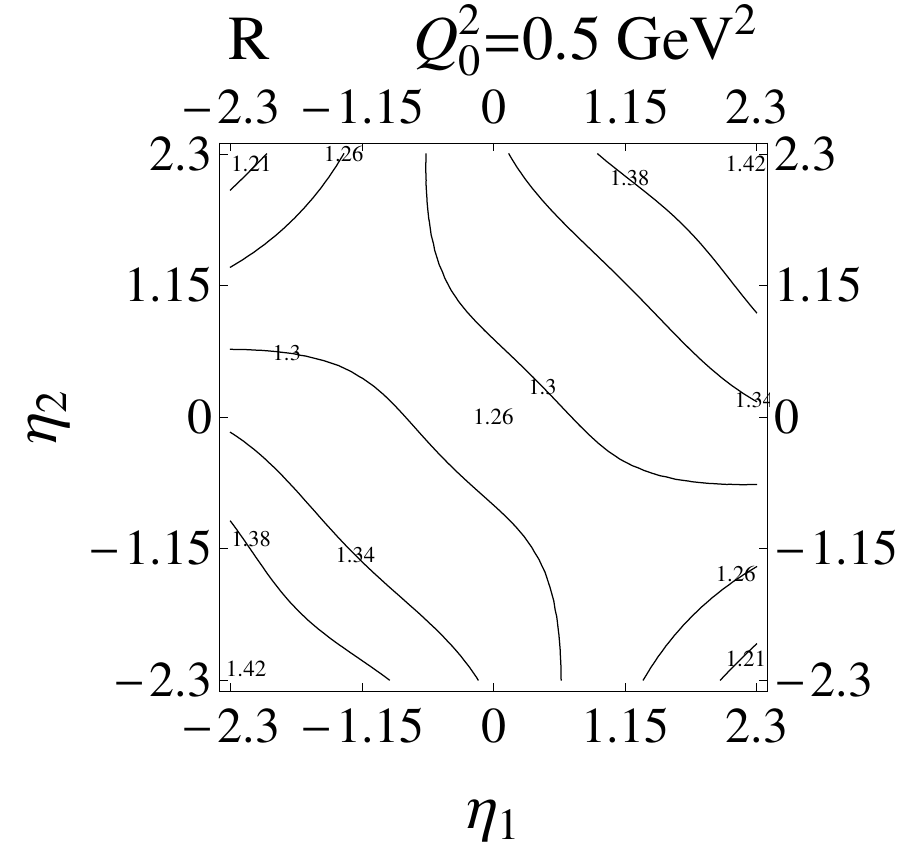}
\includegraphics[scale=0.6]{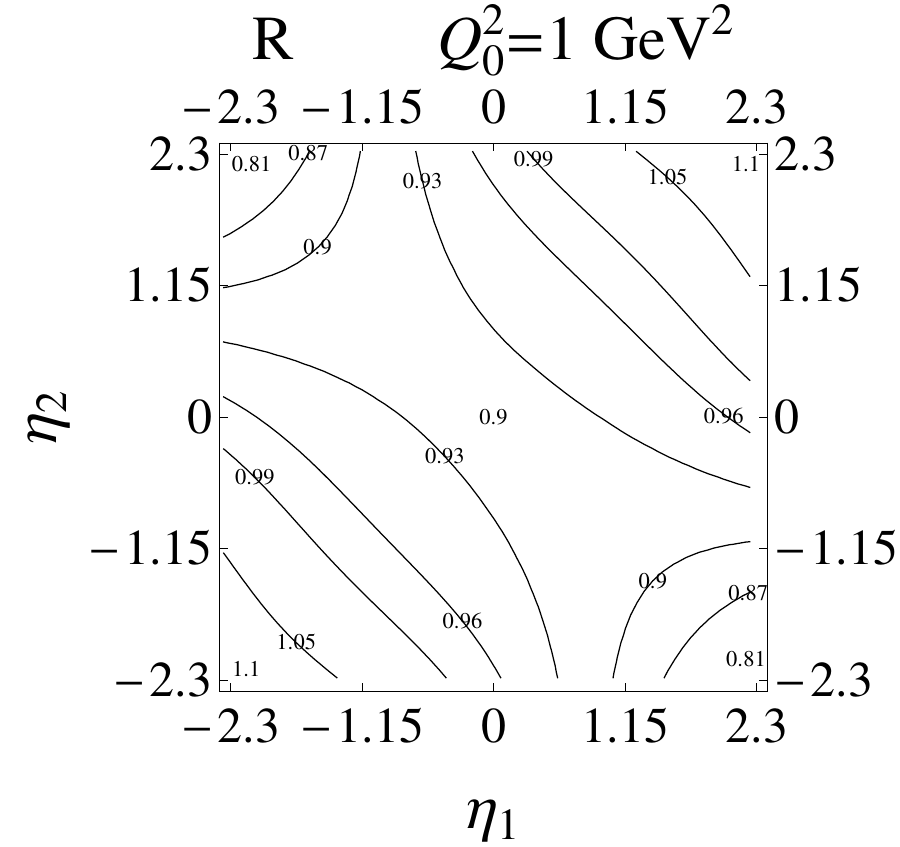}
\caption{\label{FigLHCgg05} Rapidity dependence of the $R$ factor for two pairs of $p_\perp=50\,\GeV$ jets produced in gluon-gluon collisions}
\end{center}
\end{figurehere}

\begin{figure}
\begin{center}
\includegraphics[width=0.6\textwidth]{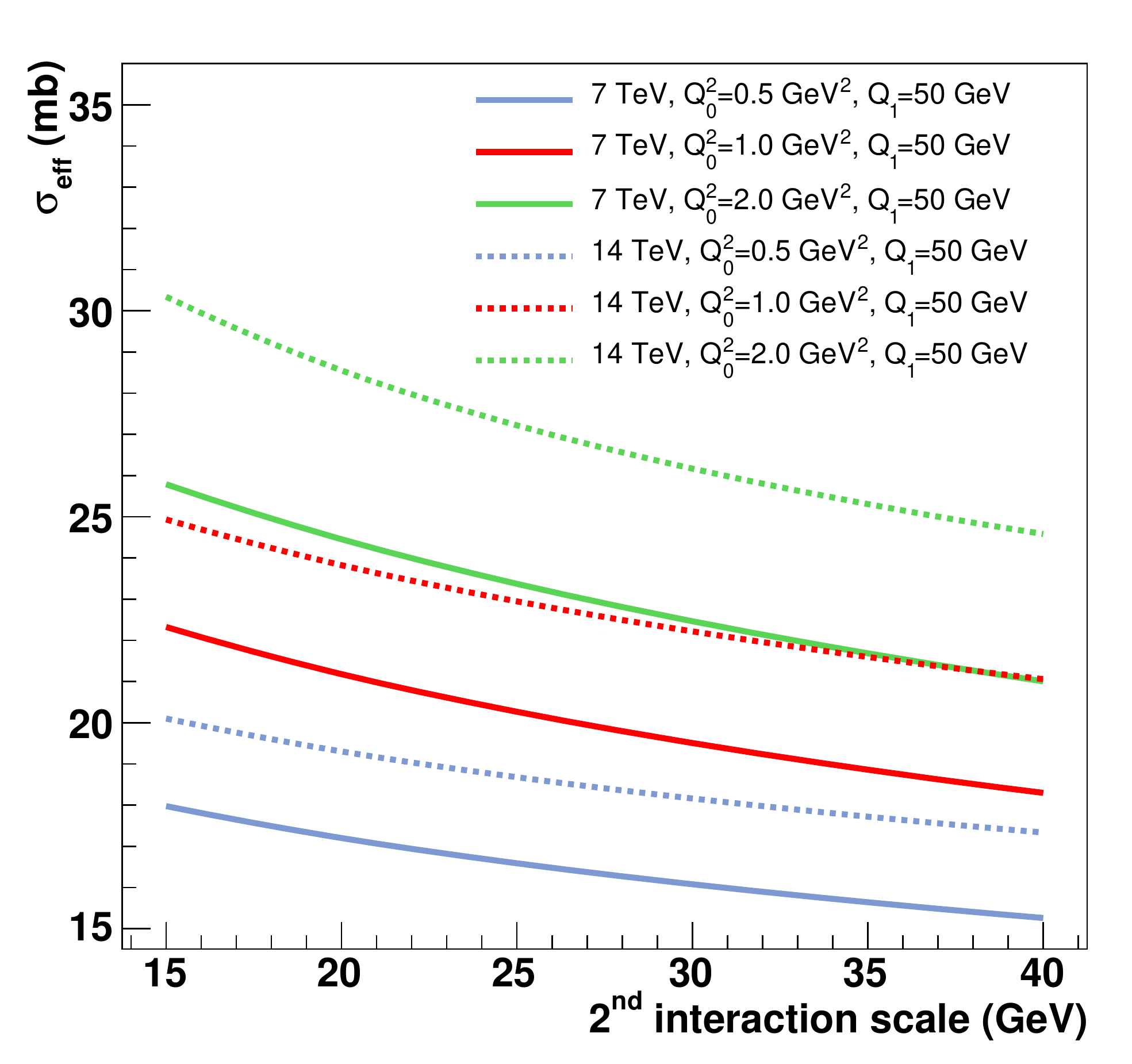}
\caption{\label{FigJET} \effs for two dijets in DPS at the LHC.}
\label{BG1}
\end{center}
\end{figure}

Dependence on the hardness parameters of the DPS process of double gluon--gluon collisions is illustrated in Fig.~\ref{FigJET}.
For the sake of illustration, we have chosen  the value of the $p_\perp$ cutoff parameter, varied
$Q_0^2=0.5,1,2\,\GeV^2$, and calculated the \effs as a function of transverse momenta of the second dijet. \cite{BG1}

 For considered $\sqrt{s}, p_\perp$  range, $R$ increases by about 15--25\%\ with increase of the hardness of one of the jet pairs. This corresponds to approximately 10\%\ drop of $\sigma_\eff$.

Finally, in Fig.~\ref{FigLHCwjj05} we show the rapidity profile of the $R$ ratio for the process of production of the vector boson,
$u\bar{d}\to W^+$, accompanied by an additional pair of (nearly back-to-back) jets with transverse momenta $p_\perp = 30\,\GeV$ produces in a gluon--gluon collision.
\begin{figurehere}
\begin{center}
\vspace{-1cm}\includegraphics[scale=0.6]{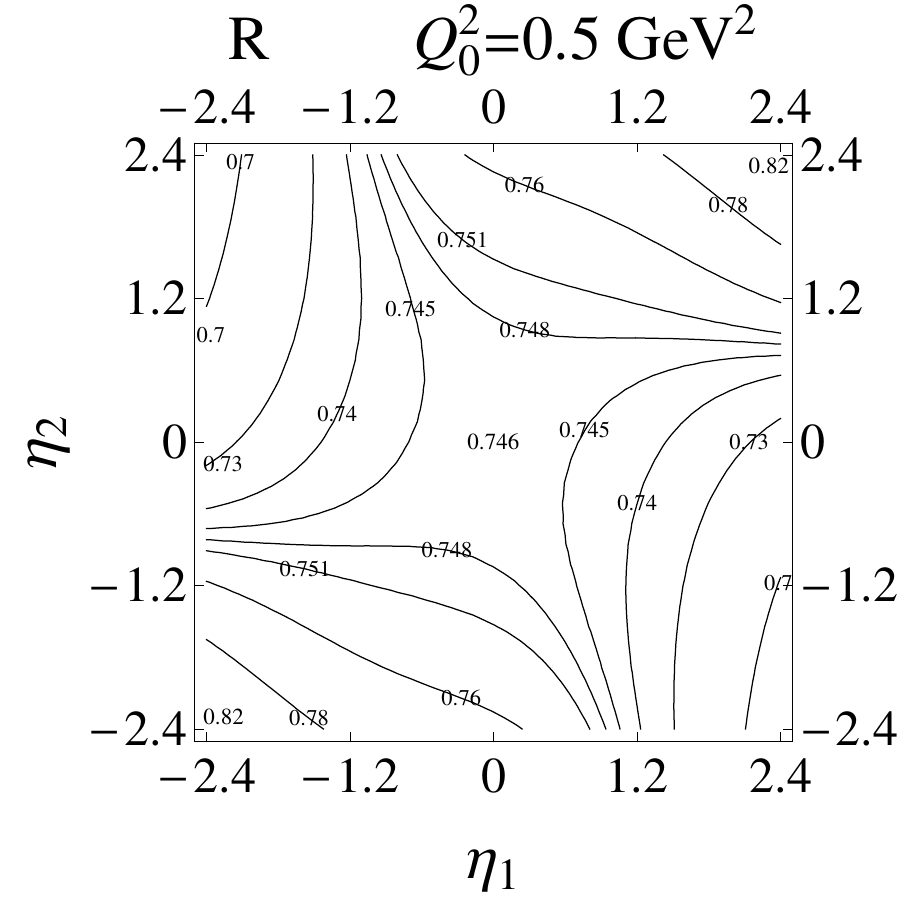}
\includegraphics[scale=0.6]{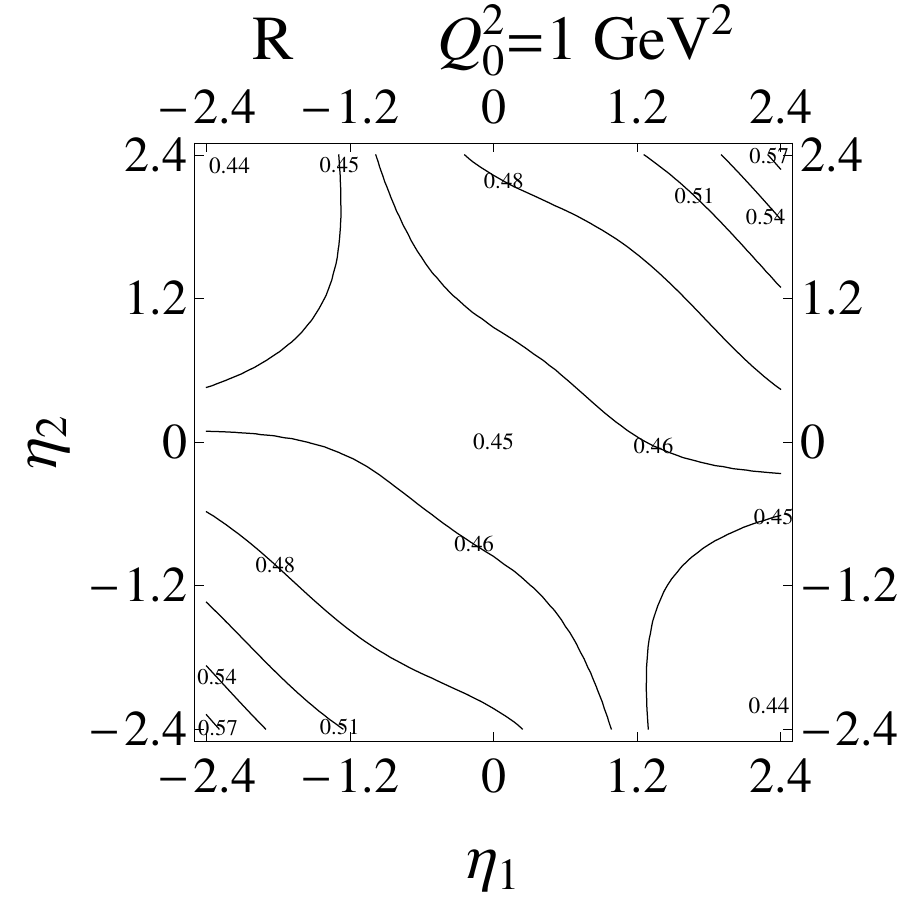}
\caption{\label{FigLHCwjj05} Ratio $R$ for production of $W$ plus a pair of $p_\perp \simeq 30\,\GeV$ gluon jets}
\end{center}
\end{figurehere}	
It is interesting to notice that the effect of perturbatively induced parton--parton correlations
is maximal for equal rapidities of the $W$ and the jet pair, and slowly diminishes  when they separate.
This feature is more pronounced when the cutoff parameter $Q_0^2$ is taken larger.
In  this case the PT correlation becomes smaller and, at the same time, exhibits a stronger rapidity dependence.

\begin{figure}[h]
  \hspace{1.5cm}\includegraphics[scale=0.35]{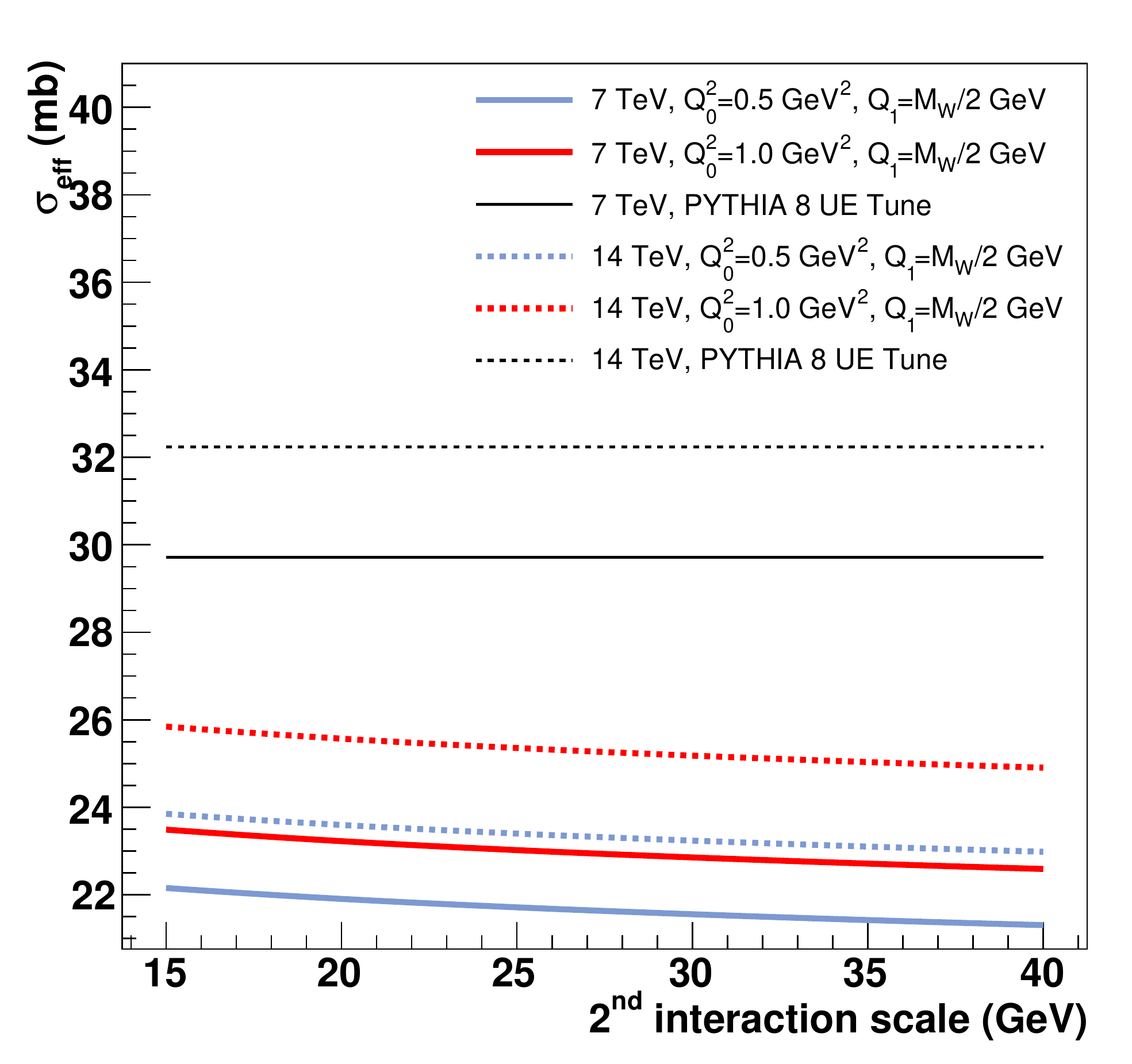}
\caption{ $\sigma_{eff}$ for Wjj processes  as function of a transverse scale of a dijet.}
   \label{FigJETW}
 \end{figure}

The recent ATLAS study \cite{Aad:2013bjm} reported for this process the value
$\sigma_{\eff} = 15 \pm 3\>^{+5}_{-3}$~mb
which is consistent with the expected enhancement due to contribution of the \12\ DPS channel, see Eq. \ref{eq:Rdef}.
The characteristic feature of our approach is that
$\sigma_{eff}$ depends  both on the  longitudinal  fractions and transverse scale. For example,  consider Wjj processes:
Fig.~\ref{FigJETW} presents the   dependence of $\sigma_{eff}$ on the transverse momenta of jets of the second pair, $ p_\perp \equiv p_{\perp2}$.

We observe that within the
considered kinematic
range, $R$ increases by about 15--25\%\ with increase of the hardness of one of the jet pairs. This corresponds to approximately 10\%\  decrease of $\sigma_\eff$.	
It is interesting to notice that the effect of perturbatively induced parton--parton correlations
is maximal for equal rapidities of the $W$ and the jet pair, and diminishes with increase of the rapidity interval
between W and 2j.
This feature is more pronounced when the cutoff parameter $Q_0^2$ is taken larger, so that the
pQCD
 correlation becomes smaller and, at the same time, exhibits a stronger rapidity dependence.

Theoretical derivation of the effective interaction area $\sigma_\eff$ (``effective cross section'') in \cite{BDFS1,BDFS2,BDFS3,BDFS4} relied on certain assumptions and approximations.
Our approach to perturbative QCD effects in DPS developed in \cite{BDFS2} was essentially probabilistic.
In particular, we did not discuss the issue of possible interference between \12\ and \22 two-parton amplitudes. One can argue that such eventuality should be strongly suppressed.
Indeed, spatial properties of accompanying radiation produced by so different configurations make them unlikely to interfere, since in the \22\ mechanism a typical transverse distance between two partons from the hadron w.f.\ is of order of the hadron size,
while in the \12\ case it is much smaller and is determined by a hard scale.
Moreover, we disregarded potential contributions from non-diagonal interference diagrams
that are due to crosstalk between partons in the amplitude and the conjugated amplitude. Such contributions appear to be  negligible in the kinematic region under consideration \cite{Gauntnew1}.

Finally, our prediction for the DPS cross sections was based on a model assumption of the absence of NP two-parton correlations in the proton.
This assumption is
the simplest guess. One routinely makes it due to the lack of any firsthand information about such correlations.
In \cite{BDFS3} we have pointed out a source of genuine non-perturbative two-parton correlations that should come into play for very small $x$ values, $x\ll 10^{-3}$, and estimated its magnitude via inelastic diffraction in the framework of the Regge--Gribov picture of high energy hadron interactions. The theory of small $x$ NP correlations and their role are discussed in the next chapter.

In order to be able to reliably extract the DPS physics, one has to learn how to theoretically predict  contribution of  two parton collision  with production of two hard systems (four jets in particular).
This is the dominant channel, and it is only in the back-to-back kinematics that the \22\ and \12\ DPS processes become competitive with it.
Among first subleading pQCD corrections to the \oo\ amplitude, there is a loop graph that looks
like a two-by-two parton collision. However this resemblance is deceptive.
 Unlike the \22\ and \12\ contributions, this specific correction does not depend on the spatial distribution of partons in the proton (information encoded in $\sigma_\eff$), it is not power enhanced in the region of small transverse momenta of hard systems, and therefore does not belong to the DPS mechanism~\cite{BDFS1,BDFS2,stirling1}.
Treating the the amplitude  corresponding  to splitting of two incoming partons at the one --lop level,
corresponds to the two-loop accuracy for the cross section.
Until this accuracy is achieved, the values of $\sigma_\eff$ extracted by experiments should be considered as tentative.

Our first conclusion is that in the kinematical region explored by the Tevatron and the LHC
experiments, the $x$-dependence of $\sigma_\eff$ turns out to be rather mild.
This by no means implies, however, that $\sigma_\eff$ can be looked upon as any sort of a universal number.
On the contrary, we see that the presence of the perturbative correlation due to the \12\ DPS mechanism results in the dependence of  $\sigma_\eff$ not only on the parton momentum fractions $x_i$ and on the hardness parameters, but also on the type of the DPS process.

For example, in the case of golden DPS channel of production of two same sign $W$ bosons \cite{stirling} the discussed mechanism leads to expectation of significantly larger $\sigma_\eff$ than for, say, $W$ plus two jets process.
Indeed, the comparison of the values of $R$ for central production of two gluon jet pairs, $Wjj$ and $W^+W^+$ (with jet transverse momenta $p_\perp\simeq M_W/2$),
gives ($\sqrt{s}= 7\,{\rm TeV} , \> \eta_1=\eta_2=0$)
\beq
\begin{array}{rl}q
R(jj+jj)  & = 1.18 \> (0.81) \\
R(W+jj) & = 0.75 \> (0.45) \\
R(W^+W^+)  & = 0.49 \> (0.26)
\end{array}
\eeq
for $
 Q_0^2=0.5\>(1.0)\, \GeV^2$.
As a result of the different
magnitude of the perturbative correlation contribution for different processes, the effective interaction areas $\sigma_\eff$ comes out to be significantly different for the three processes:
\beq
\begin{array}{rl}
jj+jj : &\quad \sigma_\eff = 14.5 \div 20 \,\mbox{mb},\\
W+jj :&\quad \sigma_\eff = 20 \div 23.5 \,\mbox{mb},\\
W^+W^+ : &\quad \sigma_\eff = 21.5 \div 25.4 \,\mbox{mb}.
\end{array}
\eeq
In all cases  the effective cross section is smaller for lower $Q_0^2$ due  to a more developed perturbative parton cascades.

In difference from
the $W^+W^+$ channel, the {\em double Drell-Yan process}\/ favors the \12\ mechanism, $g\to u\bar{u}$.
As a result, the effective interaction area in this case turns out to be significantly smaller.
For example, for the central production of two $Z$ bosons at $\sqrt{s}= 7 {\rm TeV}$ we find
\beq
R(ZZ) = 1.03 \> (0.73), \mbox{corresponding to}
  \quad \sigma_\eff (ZZ) = 15.9 \div 18.5 \,\mbox{mb}.
\eeq
\par The results for \effs for higher LHC energies are  quite close (within the accuracy of measurements), cf. Figs.~\ref{BG1}, \ref{FigJETW}, and have similar pattern.

We mentioned above that
an important feature of the \12 mechanism is its dependence on the hardness of the process.
With increase of $Q_i^2$, the \12\ to \22\ ratio $R$ is predicted to increase rather rapidly, resulting in smaller values of  $\sigma_\eff$.
At the same time, with decrease of the $p_\perp$ of the jets this contribution decreases.
We have seen above, that such a trend is consistent with the D0 data for $x \geq 10^{-2}$, Fig.~\ref{D0plot}.

\section{Non-factorized
 contribution
  to $_2D$ at the initial  $Q_0$ scale.}
\subsection{Basic ideas}
 \par There is an additional contribution to the  DPS at small $x$ which is related to the soft dynamics.
It was first discussed in \cite{BDFS3}, and in a more  detail in  \cite{BS1}.
  It was demonstrated in \cite{BS1} that soft dynamics leads to positive correlations between partons at small $x$  which have to be included in the calculation of the DPS cross section. These soft correlations can be calculated using the connection between correlation effects
  in MPI and inelastic
diffraction.  The  emerging
non-factorized contribution to $_2$GPD  is calculated at the initial scale $Q_0^2$ that separates soft and hard physics and which we consider as the starting scale for the DGLAP evolution.
  One expects that for this scale
the single parton distributions at small $x$ are given by the soft Pomeron  and soft Reggeon exchange.

 \begin{figure}[h]  
 \hspace{3.0cm}
\includegraphics[width=0.55\textwidth]{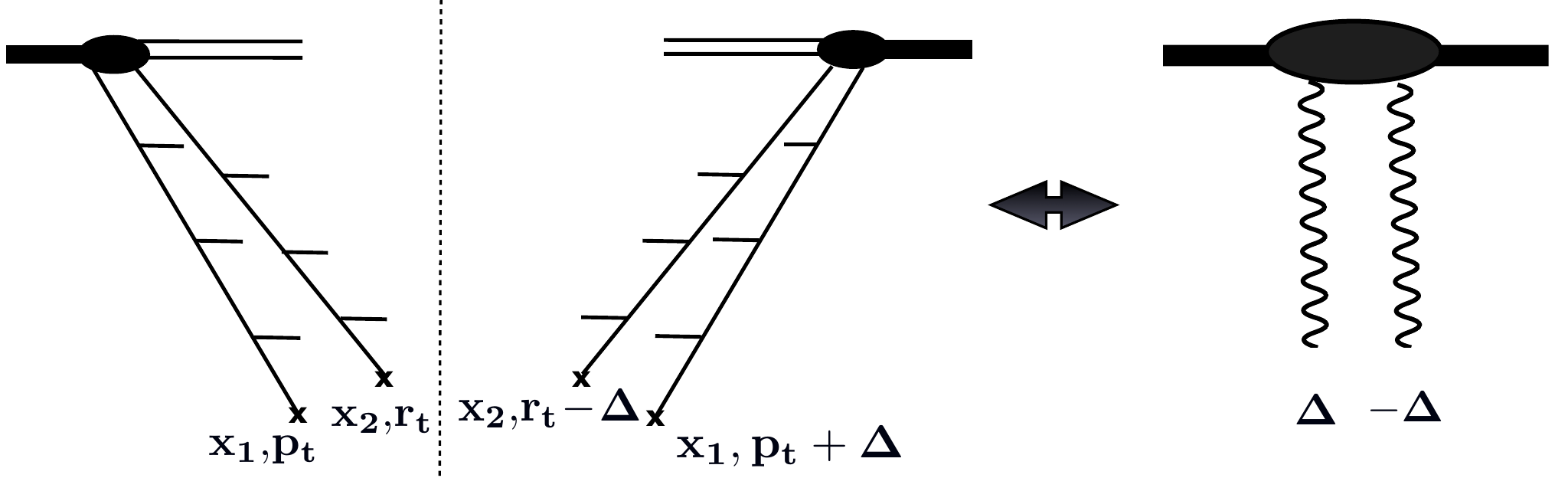}
   \caption{ $_2$GPD as a two Pomeron exchange}
    \label{geom1}
 \end{figure}

\begin{figure}[h]  
  \hspace*{2.8cm}
  \includegraphics[width=0.60\textwidth]{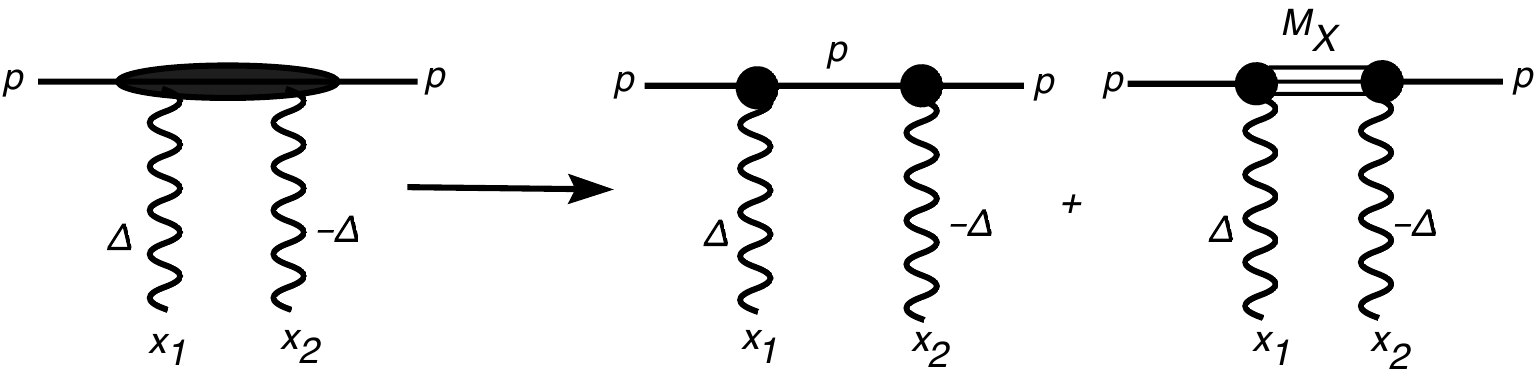}
  \vspace*{-0.3cm}
   \caption{$2\Pomeron$
   contribution to $_2$D and corresponding Reggeon diagrams}
    \label{geom2}
 \end{figure}
The diagrams of Fig.~\ref{geom1},\ref{geom2} lead to a simple expression for the non-factorizable/correlated  contribution.
(see \cite{BS1} for details).
For the correlated contribution we have,
\begin{eqnarray}
_2D(x_1,x_2,Q_0^2)_{nf}=c_{3\Pomeron}\int^1_{x_m/a} \frac{dx}{x^2}D(x_1/x,Q_0^2)D(x_2/x,Q_0^2)(\frac{1}{x})^{\alpha_{\Pomeron}}.\nonumber\\[10pt]
+c_{\Pomeron\Pomeron\R}\int^1_{x_m/a} \frac{dx}{x^2}D(x_1/x,Q_0^2)D(x_2/x,Q_0^2)(\frac{1}{x})^{\alpha_{\rm \R}}.
\label{d1}
\end{eqnarray}
Here  $x_m=max(x_1, x_2)$.  We also introduced an additional factor  of $a=0.1$ in the limit of integration over $x$ (or, equivalently, the limit of integration over diffraction masses $M^2$) to take into account that the Pomeron exchanges
occupy  at least two  units in rapidity, i.e. $M^2<0.1\cdot min(s_1,s_2)$ ($s_{1,2}=m_0^2/x_{1,2}$), or $x>max(x_1,x_2)/0.1$, where
 $m_0^2=m_N^2=1$ GeV$^2$ is the low limit of integration over diffraction masses.  Here $c_{3\Pomeron}$ and
$c_{\Pomeron\Pomeron\R}$ are normalized three Pomeron and Pomeron-Pomeron-Reggeon vertices.
We determine   $c_{3\Pomeron}$ and $c_{\Pomeron\Pomeron\R}$ from the HERA data \cite{H1}
for the   ratio of inelastic and elastic diffraction at $t=0$:
$\omega\equiv { {d\sigma_{in.\,  dif.}\over d t}\over {d\sigma_{el}\over d t}}{\left. \right\vert_{t=0} } =0.25 \pm 0.05,$
and from analysis of diffraction for large $x$ carried in \cite{Luna}, which shows that $c_{\Pomeron\Pomeron\R}\sim 1.5 c_{3\Pomeron}$
We are considering   here
 relatively low energies
 (relative large $x$)
  and a rather modest energy interval. Hence
   we
   neglect energy dependence of
   $c_{3\Pomeron}$.
    Numerically, we obtain $c_{3\Pomeron}=0.075\pm 0.015, c_{\Pomeron\Pomeron\R}\sim 0.11\pm 0.03$ for $Q_0^2=0.5 $ GeV$^2$
    and $c_{3\Pomeron }=0.08\pm 0.015$ and $c_{\Pomeron\Pomeron\R}=0.12\pm 0.03$ for $Q_0^2=1. $ GeV$^2$, using the Pomeron intercept values given below.
Note that the intercept of the Pomeron that splits into 2 (region between two  blobs in fig. 3) is always 1.1 for $t=0$, i.e. this  Pomeron
is by definition soft, and the intercept of the Reggeon is 0.5.
\par  For the parton  density in the ladder we use
 \cite{BS1}:
$xD(x,Q_0^2)=\frac{1-x}{x^{\lambda (Q_0^2)}}$,
where the small x intercept of the parton density $\lambda$ is taken from
 the GRV parametrization \cite{GRV98} for the nucleon gluon  pdf at $Q_0^2$ at small x. Numerically $\lambda(0.5 {\rm GeV}^2)\sim 0.27$,
 $\lambda(1.0 {\rm GeV}^2)\sim 0.31$.
\par Consider now the $t= - \Delta^2$ dependence of the above expressions.
The t-dependence of the  factorized contribution to $_2D_f$ is given by
\beq F(t)=F_{2g}(x_1,t)\cdot F_{2g}(x_2,t)=\exp((B_{\rm el}(x_1)+B_{\rm el}(x_2))t/2),\eeq
where $F_{2g}$ is the two gluon nucleon form factor. The t-dependence of the non-factorized term Eq.~\ref{d1}
is given by the t-dependence of the inelastic diffraction: $\exp(B_{\rm in}t)$.
Using the  exponential parameterization $\exp(B_{\rm in}t)$ for the t-dependence of the square of the {\em inelastic vertex}\/ $pM_X\!\Pomeron$,
the experimentally measured ratio of the slopes $B_{\rm in}/B_{\rm el}  \simeq 0.28$ \cite{Aaron:2009xp}
translates into the absolute value $B_{\rm in} = 1.4 \div 1.7\, {\rm GeV}^2$.
\par The evolution of the  initial conditions, Eq.
\ref{d1}, is given  by
 \begin{eqnarray}
 _2D(x_1,x_2,Q_1^2,Q_2^2)_{\rm nf}&=&\int_{x_1}^{1}\frac{dz_1}{z_1}\int_{x_2}^{1}\frac{dz_2}{z_2}G(x_1/z_1,Q_1^2,Q_0^2)G(x_2/z_2,Q_2^2,Q_0^2) \nonumber\\[10pt]
&\times& _2D(z_1,z_2,Q_0^2)_{\rm nf},
\label{r}
 \end{eqnarray}
  where $G(x_1/z_1,Q_1^2,Q_0^2)$ is the conventional DGLAP gluon-gluon kernel \cite{DDT} which
  describes evolution from $Q_0^2$ to $Q_1^2$.
In our calculation we neglect initial sea quark
 densities in the Pomeron at scale $Q_0^2$ (obviously Pomeron does not receive contribution from the valence quarks).
We refer to \cite{BS5} for numerical calculation of K.
\subsection{\effs \, in the central kinematics}.
The enhancement coefficient is now
 given by the \beq
R=R_{pQCD}+R_{soft}.
\eeq
 where $R_{pQCD}$ corresponds to the contribution of \12 pQCD mechanism (Fig.~3 right) and was calculated in \cite{BDFS4}, while
the expression for $R_{soft}$ is given by
\beq
R_{soft}=\frac{4K}{1+B_{inel}/B_{el}}+\frac{K^2B_{el}}{B_{in}}+KR_{pQCD}B_{el}/B_{inel},
\eeq
where  we calculate all factors for $x_1=x_2=x_3=x_4=\sqrt{4Q^2/s}$, with $s$ being invariant
energy of the collision.
We present our numerical results in
  Figs.~\ref{5b3},\ref{5b1}:
 \begin{figure}
\hspace{1cm}\includegraphics[scale=0.63]{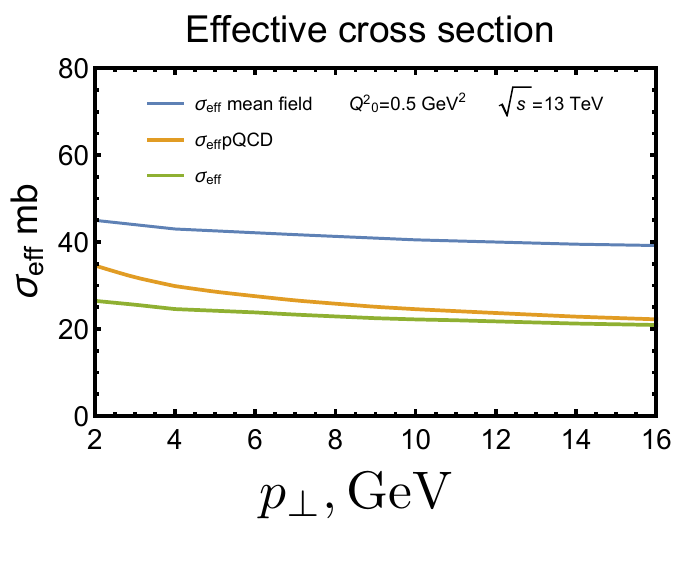}
\includegraphics[scale=0.68]{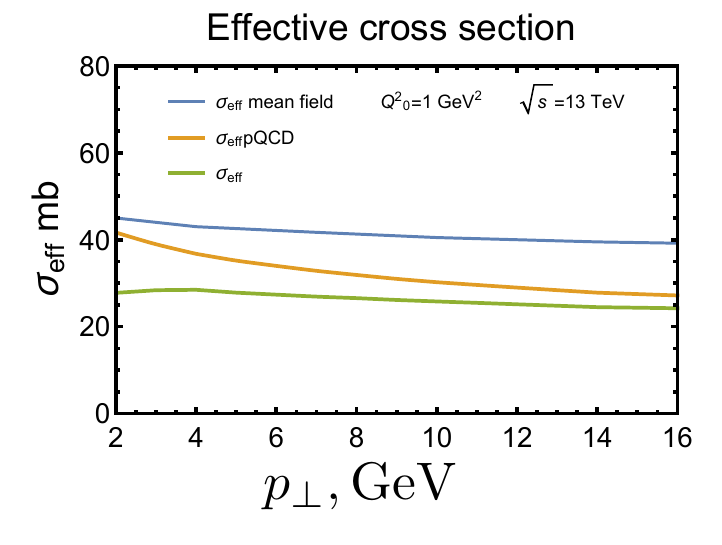}
\caption{\effs \,  as a function of the  transverse scale $p_\perp$ for $Q_0^2=0.5$ (left),and $Q_0^2=1$ GeV$^2$  (right) in the central kinematics.
We present the mean field, the mean field plus \12 mechanism and total \effs  for $\sqrt{s} $=  13 TeV.}
\label{5b3}
\end{figure}

\begin{figurehere}
\includegraphics[scale=0.64]{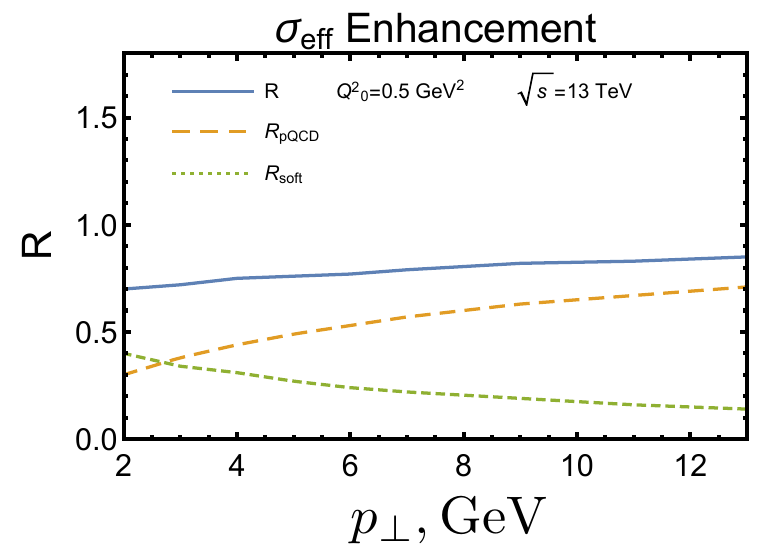}
\includegraphics[scale=0.7]{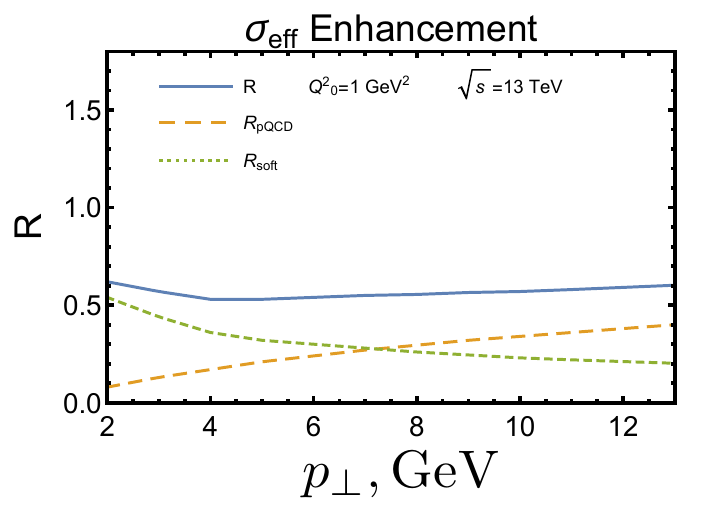}
\caption{R for different $Q_0^2$ and $\sqrt{s} $ =13 TeV.}
\label{5b1}
\end{figurehere}

\par In addition, in order to
illustrate
 the
 picture of \effs behavior in both UE and DPS, in
  Fig.~\ref{5b3} we give the example of the $p_\perp$ dependence
of \effs
for the transverse momenta region 2--50 GeV for $Q_0^2$=0.5 GeV$^2$ (for $Q_0^2$=1 GeV$^2$
the behavior is very similar).
\begin{figure}
\includegraphics[scale=0.83]{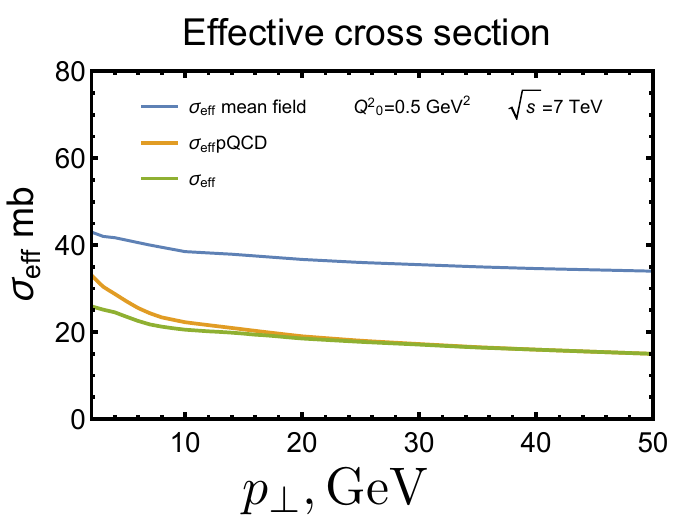}
\includegraphics[scale=0.6]{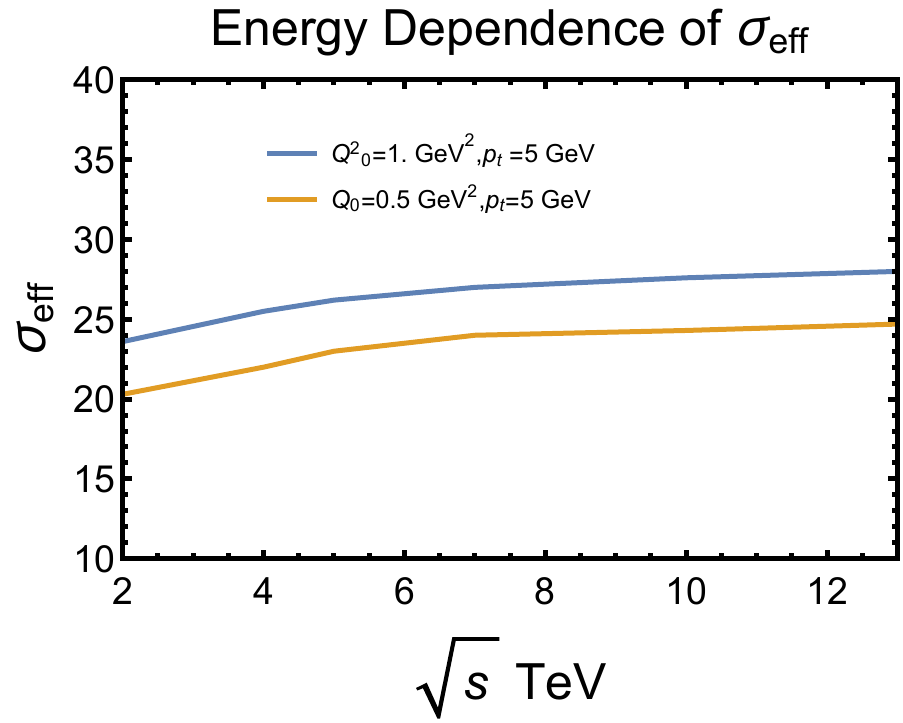}
\caption{left:\effs for the  entire transverse momenta region ($Q_0^2=0.5$ GeV$^2$),right:The characteristic energy dependence of \effs on c.m.s. energy $\sqrt{s}$.}
\label{5b7}
\end{figure}
\par We also studied the energy dependence of \effs for fixed transverse momenta $p_\perp$ on $s$  in the UE kinematic region in the energy region from Tevatron to LHC. We find that \effs slowly increases with $s$ and practically 
flattens out
at the top of the LHC energies, see  Fig.~\ref{5b7} right.

 In order to understand
the evolution of \effs at higher incident   energies for given transverse scale we would  need the information on the $x$ dependence of the two-gluon form factor
for small x$\le 10^{-4}$ and of the inelastic diffraction which are likely to come from the current analyses of the $J/\psi$ diffractive  production in the ultraperipheral collisions at the LHC.

Our current estimates of non-factorizable contribution should be considered as
semiquantitative due to the
 large uncertainties in diffraction parameters as well as the use of the
"effective" values for the reggeon/pomeron parameters (which  include
screening corrections very roughly). Nevertheless,
our results indicate
a number of basic features of soft nonperturbative parton - parton correlations which are relevant for the
 central LHC dynamics.
\par (i)  For large transverse momenta, relevant for hard DPS scattering, soft effects are small and essentially negligible, contributing only $5\%$ to the enhancement coefficient $R$ if we start evolution at  the scale $Q_0^2=0.5$ GeV$^2$, and $10-15\%$ for the starting scale of  1 GeV$^2$, for $p_\perp \sim 15--20$ GeV.
Thus they do not influence detailed hard DPS studies described  in the previous sections.
Our results also indicate that the characteristic transverse momentum $p_{t0}$, for  which soft correlations constitute given fixed
fraction  of the enhancement factor $R$ rapidly increase with $s$.
\par (ii)  The soft non-factorizable contributions may contribute significantly in the underlying event dynamics, especially
at the scales $p_{\perp}=2\div 4$ GeV where they are responsible for about 50\% of
the difference between mean field result and full prediction for \effs
 for $Q_0^2 \mbox{=0.5 GeV}^2$ case. If we would start evolution at
$Q_0^2 =\mbox{1 GeV}^2$, soft effects  would dominate up to scale $p_{\perp} \sim $4 GeV.
In the UE the account of the soft contribution  leads to stabilization of the results for \effs,  and to its slower  decrease with increase of $p_\perp$ than in the approximation in which only perturbative correlations, i.e. the \12 mechanism is included.
  These values for \effs for UE, especially for scales 2--4 GeV are very close to the ones used by Pythia.

 We see that the new framework   gives a  reasonable  description of the data over the full transverse momenta range, with weaker  dependence of the quality of the fit on the starting point of the evolution $Q_0$ than in \cite{BG1}.
\par (iii) The evolution of \effs with transverse scale is stabilized for the UE regime, as shown in
Fig.~\ref{5b3}
 leading to an almost plateau like behavior  with
a slight decrease with increase of the  transverse scale.
\par  (iv) The inclusion of the soft correlations  stabilizes the incident energy dependence of  \effs. It changes only slightly  between 3.5 TeV and 6.5 TeV proton collision energies  for the same transverse scale for small $p_\perp$. In other words,  the increase of the soft correlations compensates the decrease of the relative pQCD contribution with an increase of the collision energy due to  decrease of effective $x_i$.
\par We refer the reader to the original papers \cite{BS4,BS5} for more details.

One of the processes which is sensitive to the non-factorizable contribution is production of  double open charm in the forward kinematics which was recently studied  in the LHCb experiment. We find that the mean field approximation for the double parton GPD, which neglects parton - parton correlations, underestimates the observed rate by a factor of two. The enhancement due to the perturbative QCD correlation \12 mechanism which explains the rate of double parton interactions at the central rapidities is found to explain 60 $\div$ 80 \% of the discrepancy. We find \cite{BS4} that  non-factorized  contributions  to the initial conditions for the DGLAP collinear evolution of the double parton GPD discussed above play an important role in this kinematics. Combined, the two correlation mechanisms provide a good description of the rate of double charm production reported by the LHCb\cite{LHCb} with the result weakly sensitive to   to the starting point of the QCD evolution.  At the same time we cannot reproduce small values of \effs for the double $J/\psi $ channel reported by the LHCb which may indicate a more complicated pQCD of charmonium production.

\section{MPI in proton - nucleus scattering}
Above we considered the case of $pp$ scattering. The same formalism is applicable for collision of any two hadrons and in particular for the proton - nucleus scattering.  Cross sections of the MPI processes for the $pA$ case were  first calculated in the parton model approximation in \cite{ST}. It was demonstrated that the double, triple,... MPI are  strongly enhanced in the proton collisions with heavy nuclei due to a possibility of hard collisions occurring simultaneously on 2 (3,..) nucleons at the same impact parameter. It was also   emphasized that a comparison of the MPI in $pp$ and $pA$ scattering would allow to  study longitudinal correlations of partons in nucleons \cite{ST}. Further, it was demonstrated in \cite{BS2} that corrections to the parton model expression in the mean field approximation are much smaller than in the $pp$ case.

\begin{figure}[t]
\hspace{2cm}\includegraphics[width=.70\textwidth]{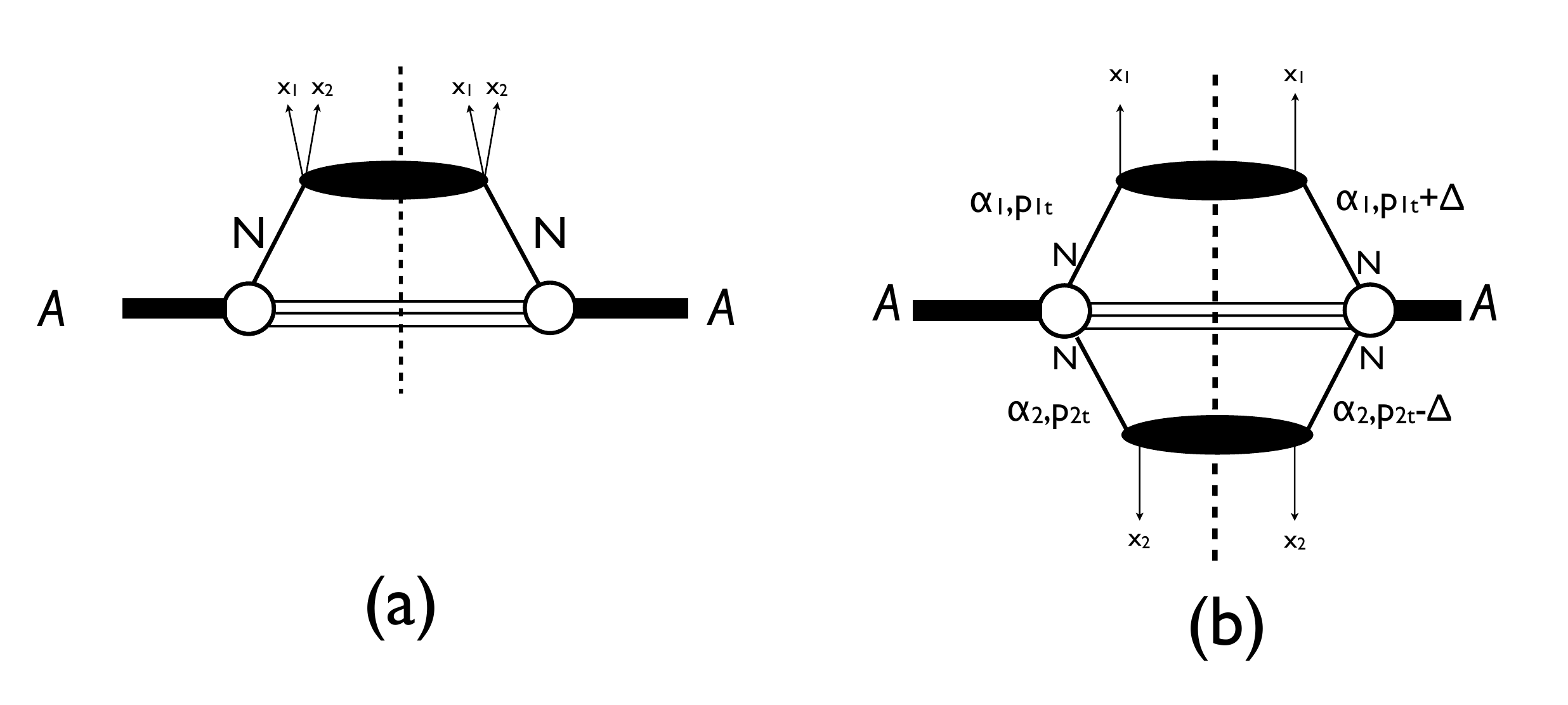}
\caption[]{Impulse approximation and two nucleon contributions to DPS in $pA$ scattering }
\label{pafig}
\end{figure}

The technique we described in section 2 allows to perform the calculation in a very compact form \cite{BS2}. We will focus on the case of DPS. In this case we have two contributions - the impulse approximation, corresponding to two partons of the nucleus involved in the collision belonging to the same nucleon (Fig.~\ref{pafig}a), $\sigma_1$:
\begin{equation}
\sigma_1= A \sigma_{NN}
\end{equation}
and two  different nucleons, $\sigma_2$ (Fig.~\ref{pafig}b). Since the b-dependence of the nuclear density is much slower than that for the nucleon $\sigma_2$ is not sensitive to the transverse distance between the partons of the nucleon and hence proportional to the double parton distribution.
Picking up two partons at similar impact parameters in the heavy nucleus is $\propto A^{4/3}$, leading to
 $\sigma_2/\sigma_1 \propto A^{1/3}$

For simplicity we restrict the discussion to the case of $x_i^{(A)}\ge 0.01$ where interference effects corresponding to $x_1^{(A)}  (x_2^{(A)}) $ belongs to one nucleon in the $\left| in \right> $ state and to
another nucleon in the $\left< out \right| $ state. For analysis  of the effect of the leading twist nuclear  shadowing see \cite{BSpd}.

Similar to the $pp$ case, the  expression for $\sigma_2$ is proportion to the integral over $\Delta$ of the product of the double GPDs of proton and nucleus.
The two GPD form factor of the nucleus is
\begin{equation}
_2GPD_A(\Delta)=F_A(\Delta,-\Delta)\cdot F_g^2(\Delta,x).
\end{equation}
Here $F_A(\Delta,-\Delta)$ is the two body nuclear form factor.  In the mean field approximation for the nucleus wave function:
\begin{equation}
F_A(\Delta,-\Delta)= A^2F_A^2(\Delta)\approx A^2\exp(-R_A^2 \Delta^2/3),
\end{equation}
where  $F_A(\Delta)$ is the nucleus single   body form factor normalized to one at $\Delta=0$.
Since the $\Delta^2$ dependence of $F_A^2(\Delta)$ is much stronger than that of $F_g^4(\Delta, x)$ the
later can be neglected in the integral
\begin{equation}
\int  d^2 \Delta  _2GPD(\Delta,N)\cdot _2GPD(\Delta,A)= \int   d^2  A^2\Delta F_A^2(\Delta).
\end{equation}
It is convenient at this point to switch to the impact parameter representation using $\int  d^2 \Delta A^2F_A^2(\Delta)= \int d^2b T_A^2(b)$, where $T_A(b) $ is
 is the nuclear thickness function normalized to A:
 $\int d^2b T_A(b) = A$ leading to
\begin{eqnarray}
\frac{\sigma_4(x_1',x_2',x_1,x_2)}{d\hat t_1d\hat t_2} =
\sigma_4(NN) \cdot
\underbrace{\int T(b)d^2b}_{ A} + \nonumber\\[10pt]
+
 {f_p(x_1', x_2') \over f_p(x_1') f_p(x_2')}
\frac{d\sigma_{2}(x_1',x_1)}{d\hat t_1}\,
\frac{d\sigma_{2}(x_2',x_2)}{d\hat t_2}
\underbrace{\int T^2(b)d^2b}_{\propto A^{4/3}}\, .
\label{doubleq}
\end{eqnarray}
 In the mean field approximation for NN scattering
we use in this article, the \12\  contribution is strongly suppressed in $pA$ scattering:
 $R_{pA}/ R_{NN}\approx 1/5$ \cite{BS2}.

 We find for the ratio double scattering and single scattering terms:
 \begin{equation}
 \sigma_2/\sigma_1= 1.1 \left({\sigma_{eff}\over \mbox{15 mb}}\right)
 \cdot \left({A\over 40}\right)^{.39} (1+R_{pp}/5),
 \end{equation}
 for $A\ge 40$.
 Hence for the typical hard kinematics: $\sigma_{eff} \sim 15 \div 20 \mbox{mb}$, $R_{NN}$ =0.8 we find the enhancement of DPS as compared   to the impulse approximation result:
 \begin{equation}
1+ \sigma_2/\sigma_1 \approx 3.5 \div 4.2.
\end{equation}
for the lead nucleus.
 Note that
 Eq. \ref{doubleq} is valid  for fixed $b$ as well. Hence we expect that for collisions with heavy nuclei at small $b$  the ratio $  \sigma_2/\sigma_1$ is enhanced by and additional  factor of $\sim 1.5$.    Hence one can look for the effect of MPI in $pA$ scattering by comparing   central and peripheral $pA$ collisions.

  Much larger enhancement is expected for higher order MPI \cite{ST}. However only chance to observe such rare events would be for rather  small $p_\perp$ and small $x_A$ where leading twist shadowing effects would significantly reduce elementary cross sections.

\section{Soft -- hard interplay in $pp$ collisions at the LHC}
 \subsection{Underlying event and transverse geometry}
The  $pp$ LHC data already provide important tests of the  transverse geometry of $pp$ collisions described in Sect.\ref{transverse}.

 Let us first consider production of a  hadron (minijet) with momentum $p_\perp$. The observable of interest
here is the transverse multiplicity, defined as the multiplicity
of particles with transverse momenta in a certain angular region
perpendicular to the transverse momentum of the trigger particle or jet
(the standard choice is the interval $60^\circ < |\Delta \phi| < 120^\circ$
relative to the jet axis; see Ref.~\cite{Affolder:2001xt} for an
illustration and discussion of the experimental definition).
In the central collisions one expects a much larger transverse multiplicity due to the presence of multiple hard and soft interactions.
At the same time
the enhancement
should be a weak function of $p_\perp$ in the region where main contribution is given by the hard mechanism \cite{FSW,Frankfurt:2010ea}. The predicted increase and eventual flattening of the
transverse multiplicity agrees well with the pattern observed in the
existing data. At $\surd s = 0.9\, \textrm{TeV}$ the transition occurs
approximately at
$p_{T, {\rm crit}} \approx 4\, \textrm{GeV}$,
at $\surd s = 1.8\, \textrm{TeV}$ at
$p_{T, {\rm crit}} \approx 5\, \textrm{GeV}$,
and  at $p_{T, {\rm crit}} = 6-8\, \textrm{GeV}$  for $7\, \textrm{TeV}$ \cite{Chatrchyan:2011id,Aad:2011qe}.
Note also that $p_{T, {\rm crit}}$ is smaller for the single hadron trigger than for a jet trigger since the leading hadron
carries a fraction $\sim 0.6 \div 0.7$ of the jet momentum,
see comparison the CMS jet data and ALICE single hadron data in Fig.~3 of  \cite{Azarkin}.

One possible interpretation is  the minimum $p_\perp$
at which  particle production due to hard collisions starts to dominate  significantly  increases with the collision energy. Another is that for the  small $p_\perp$   one selects
events with fewer DMS collisions due to cutoff on minimum $p_\perp$ which becomes stronger with increase of the incident energy. Both these effects
are  likely to be related the onset of the high gluon density regime in the central $pp$ interactions since with an increase of incident energy leads to partons in the central $pp$ collisions propagating  through stronger and stronger gluon fields.

Many further tests of the discussed picture were suggested in Ref.~\cite{Frankfurt:2010ea}. They include
(i) Check that the transverse multiplicity  does not depend
 on rapidities of the jets, (ii) Study of the  multiplicity  at
$y < 0$ for events with   jets at $y_1 \sim  y_2 \sim 2$. This would allow to check that the     transverse  multiplicity is universal  and that multiplicity in the   away and the  towards regions is similar to the  transverse multiplicity for $ y \le 0$. (iii) Studying whether transverse multiplicity is the same for quark and  gluon induced jets. Since the gluon radiation for production of  $ W^{\pm},Z$ is smaller than for the gluon dijets, a  subtraction of the radiation effect mentioned below is very important for such  comparisons.

Note that the contribution of the jet fragmentation to  the transverse cone as defined in the experimental analyses  is small but not negligible especially at smaller energies
($\sqrt{s}=0.9 \mbox{TeV}$). It would be desirable to use a more narrow transverse cone, or subtract the contribution of the jets fragmentation. Indeed,   the color flow  contribution \cite{Dokshitzer:1991wu} leads to a small residual increase of the transverse multiplicity  with $p_\perp$.
However the jet fragmentation  effect depends on $p_T$ rather than on $\sqrt{s}$. Hence it does not contribute to the growth of the transverse multiplicity, which is   a factor of $\sim 2$ between $\sqrt{s}=0.9 \mbox{TeV}$ and $\sqrt{s}=7.0 \,\mbox{TeV}$.  In fact, a subtraction of the  jet fragmentation contribution would somewhat increase the rate of the increase of the   transverse  multiplicity in the discussed energy interval. This allows to obtain the lower limit for the rate of the increase of the multiplicity in the central ($\left<b\right >\sim$ 0.6 fm) $pp$ collisions of $s^{0.17}$. It is    a bit  faster than the $s$ dependence of multiplicity in the central heavy ion collisions.

 \subsection{Correlation of soft and hard multiplicities}

 It was demonstrate recently \cite{Azarkin}  that  the rates of different hard processes observed in
jet production by CMS and in $J/\psi$, D-meson production by  ALICE
normalized to the average hard process multiplicity, $R$
 universally depend on the underlying event charged-particle multiplicity normalized to the average charged-particle multiplicity
at least until   it  becomes  four  times  higher  than  average. Note here that the recoil jet multiplicity has to be subtracted from the underlying multiplicity.

It is worth emphasizing here, that similarity between R in the CMS and ALICE  measurements is
highly non-trivial as the rapidity intervals used for determination of
$N_{ch}$  differ by a factor of $\sim$ 3.

The  ratio of the inclusive rate of hard signals at fixed $b$
to the average one in bulk events is given as follows \cite{Strikman:2011ar}
\beq
R(b) = P_2(b)\sigma_{inel}.
\label{pb0}
\eeq
 The median of the distribution over $N_{ch}$
 should  roughly correspond to the median of the distribution over impact parameters.
 For the studied inelastic sample $\sigma_{inel} \approx $55 mb. Using parametrization of $P_{2b}$ from the Appendix we find $R_{median} \approx 2$ which agrees well with the data.

 The relation Eq. \ref{pb0} breaks down when the
 multiplicity selection starts to select $b\sim 0$ corresponding to $R(0)\approx 4$.

 For $b\sim 0$ the trigger on high multiplicity starts to select configurations in colliding nucleons with larger than average number of hard collisions,  corresponding to $R(0)$.

 In this limit large fraction of the total multiplicity originates from gluon emission in processes associated with minijet production. So one can expect that in this limit  $N_{ch}/ \left< N_{ch}\right> $
 is proportional to the number of the hard collisions, $N$, and leading to the linear dependence between $N$ and   $N_{ch}/ \left< N_{ch}\right> $. This expectation is consistent with the data.

 An interesting question is whether high multiplicity events originate  from tail of distribution over number of hard collisions at $b\sim 0$ or from some correlated configurations.
  In  \cite{Strikman:2011ar} it was suggested that for the highest observed  multiplicities (which occur with probability $\sim 10^{-4} \div  10^{-5} $ fluctuations of the gluon density are important.

  \subsection{Unitarity and consistency
in multiple hard collisions}
 One of the important observations of the MC models is that to reproduce the data one needs to suppress production of minijets. PYTHIA \cite{Sjostrand:2014zea} introduces the energy dependent suppression factor
   \beq
    R(p_\perp)=p_T^4/ (p_\perp^2+p_0^2(s))^2,
 \ee
    with $p_0(\sqrt{s}=7 TeV)\approx 3 GeV/c$, corresponding to $R(p_T=4 GeV/c)=0.4$. In HERWIG \cite{Herwig} a cutoff of similar magnitude is introduced of the form $\theta(p_\perp-p_0'(s))$.

 A complimentary way to see that a mechanism of the suppression has to exist follows  from the analysis of the  restrictions related to the value of the total inelastic cross section at a fixed impact parameter \cite{Rogers:2008ua,Rogers:2009ke}.
 (Note here that
 the large inclusive cross section of production of minijets which exceeds the total inelastic cross section does not violate the $S$-channel unitarity  since it effectively measures
 multiplicity of minijet production.)

 It is possible to rewrite the cross section of the production of minijets as a series of  positive
 terms $\sigma_i = \int d^2b Pf_i(b)$,  where $Pf_i(b)$ is the probability that in the collision at fixed $b$  exactly $i$ minijet pairs are produced.
 The total probability of inelastic interaction at given $b$ is expressed through the elastic scattering amplitude (Eq. \ref{P_in_def}).

 Unitarity in the $b$ space leads to the condition
 \beq
 P_{hard}(b)= \sum_{i=1}^{\infty} Pf_i(b)\le P_{\mbox{in}} (s, b).
 \label{unitin}
 \eeq
 As we discussed in section \ref{transverse}
 distribution over $b$ for generic inelastic collisions is much broader than for hard binary collisions and that of binary collisions is much broader than of MPI events (Fig. \ref{scales}).
 As a result in a MPI model without nonperturbative correlations one finds that  for $ b\ge $1.2 fm inclusive cross section for production of minijets at given $b$ and
  $P_{hard}(b)$
  practically coincide and hence the analysis does not depend on the details of modeling.

 Numerical studies  indicate  that to satisfy inequality Eq.
 \ref{unitin} one needs to suppress production of minijets in the momentum range similar to that  introduced in the Pythia
 \cite{Sjostrand:2014zea}
 and HERWIG  \cite{Herwig} models.

 In fact one may need an even stronger cutoff. Indeed
 in Eq.\ref{unitin} we did not
 take into account that inelastic diffraction contributes a significant fraction, $\sim  15 \div 20 \%$, of $\sigma_{in}$ at the LHC and it is predominantly due to events with no minijet production (remember that even in DIS where absorptive effects are small diffraction constitutes a small fraction of the small $x$ cross section ($\le 20\%$).
 Since  for small $b$ interaction is essentially black and hence diffraction is impossible, the main contribution of diffraction to $P_{\mbox{in}}$ should concentrate at $b\ge \mbox{1.2 fm}$, leading to a need for even stronger cutoff.

 A dynamical mechanism for a  strong cutoff for the interaction at large impact parameters is not clear. Indeed, typical $x_1,x_2$ for hard collisions are $10^{-2}\div 10^{-3}$ for which pQCD work well at HERA. Also, colliding partons of the nucleon "1" ("2") propagate typically though much smaller   gluon densities of the
 the nucleon "2" ("1") so one would expect a very strong dependence of the cutoff on impact parameter. Alternatively, one would have to introduce very strong correlations for partons at the nucleon periphery.

 Understanding the origin of this phenomenon is one of the challenges for the future studies.

\section{Conclusions}
We developed the momentum space technique for describing MPI based on introduction of double parton (triple ...) GPDs.  It allows effectively introduce both the mean field approximation which is constrained by the data on single parton GPDs and developed the framework for including perturbative and nonperturbative correlations between the partons. We find that perturbative correlations enhance the  high $p_{\perp}$ DPS rates bringing into a fair agreement with most of the experiments. In the underlying event kinematics
an additional NP mechanism becomes significant which strength was estimated based on information about  double Pomeron exchange. NP mechanism  largely  compensates the increase of
\effs expected in the mean field approximation due to the increase of the gluon distribution radius with decrease of $x$.

\par Taken together these three mechanisms provide a  good description of experimental data on MPI in the entire kinematical domain, including forward heavy flavor production observed in LHCb \cite{Belyaev}.

The dijet production and even more so MPI occur at   smaller impact parameters than the soft interactions giving  leading to explanation of some of regularities of UE and to  a conclusion that  a strong suppression of minijet production even in peripheral collisions is necessary for explaining $pp$ data.

Further studies are necessary in order to  go beyond the leading log approximation as well as  to understand dynamical mechanism of the suppression of the minijet production. It would be desirable to find a way to distinguish the scenario presented here with a low $Q^2$ scale starting point  of the pQCD evolution and weak NP correlations at $x > 10^{-3}$ and a scenario where the NP correlations
are present at $x\sim 10^{-2}$ while pQCD evolution starts at significantly higher scale.
\par Next, further work may be needed to describe recent experimental data on $J/\Psi$ production in central kinematics \cite{cmsnew,lansberg}.
\par Here we focused on $x_i < 0.1 $ domain.  Large $x$ region is certainly of much interest for understanding the nucleon structure.
For example,    strong  quark�antiquark  correlations  may  arise \cite{Schweitzer:2012hh} from  dynamical  chiral
symmetry breaking. Also,  one expects significant  correlations between valence quarks.
 They could be studied in  the forward DPS for example in the production of two forward pions \cite{Strikman:2010bg}.
\section*{Acknowledgements}
We thank our coauthors M.Azarkin,  Yu.Dokshitzer, L. Frankfurt, P. Gunnellini, T. Rogers, A. Stasto, D.Treleani,
 C. Weiss, U. Wiedemann  for numerous discussions and insights. We thank V.Belyaev for discussions of the LHCb charm data.
M.S.'s  research  was  supported  by  the  US  Department  of  Energy  Office  of  Science,  Office  of
Nuclear Physics under Award No.
 DE-FG02-93ER40771.

\section*{Appendix}
The QCD factorization theorem for exclusive vector meson (VM) production\cite{Collins:1996fb}  states that in the leading twist approximation  the differential cross section of the process  $\gamma^*_L + p \to VM + p$ is given  by the  convolution of the hard block, meson wave function and generalized gluon parton distribution, $g(x_1,x_2, t\left|\right. Q^2) $,
 where $x_1, x_2$  are the
longitudinal momentum fractions of the emitted and absorbed gluon
  (we discuss here only the case of small x which is of relevance for the LHC kinematics).  Of particular interest
is the generalized parton distribution (GPD) in the ``diagonal'' case, $g(x, t | Q^2)$, where $x_1=x_2$ and denoted by $x$, and the momentum transfer to the
nucleon is in the transverse direction, with
$t = -\Delta_\perp^2$ (we follow the notation of
Refs.~\cite{FSW,Frankfurt:2010ea}). This function reduces to the
usual gluon density in the nucleon in the limit of zero momentum
transfer, $g(x, t = 0| Q^2) = g(x| Q^2)$. Its two-dimensional
Fourier transform
\begin{equation}
g(x, \rho | Q^2) \; \equiv \; \int\!\frac{d^2 \Delta_\perp}{(2 \pi)^2}
\; e^{i (\bm{\Delta}_\perp {\bm\rho })}
\; g(x, t = -{\bm{\Delta}}_\perp^2 | Q^2)
\label{gpdrho_def}
\end{equation}
describes the one--body density of gluons with given $x$ in transverse space,
with $\rho \equiv |\bm{\rho}|$ measuring the distance from the
transverse center--of--momentum of the nucleon, and is normalized
such that $\int d^2\rho \, g(x, \rho | Q^2) \;\; = \;\; g(x|Q^2). $
It is convenient to separate the information on the total
density of gluons from their spatial distribution and parametrize
the GPD in the form
\beq
g(x, t | Q^2) \;\; = \;\; g(x | Q^2) \; F_{2g}(x, t | Q^2) ,
\eeq
where the latter function satisfies $F_{2g}(x, t =0| Q^2) = 1$ and is known as
the two--gluon form factor of the nucleon. Its Fourier transform describes
the normalized spatial distribution of gluons with given $x$,
\beq
F_{2g} (x, \rho | Q^2) \; \equiv \; \int\!\frac{d^2 \Delta_\perp}{(2 \pi)^2}
\; e^{i (\bm{\Delta}_\perp \bm{\rho})}
\; F_{2g} (x, t = -{\bf{\Delta}}_\perp^2 | Q^2) ,
\label{rhoprof_def}
\eeq
with $\int d^2\rho \, F_{2g} (x, \rho | Q^2) = 1$ for any $x$.

The QCD factorization theorem predicts that the t-dependence of the VM production should be a universal function of $t$ for fixed $x$
(up to small DGLAP evolution effects).
Indeed the t-slope of the $J/\psi$ production is  practically $Q^2$ independent,
while the t-slope of the production light vector mesons approaches that of $J/\psi$ for large
$Q^2$.
The $t$--dependence of the measured differential cross sections of
exclusive processes at $|t| < 1 \, \mbox{GeV}^2$ is commonly
described either by an exponential, or by a dipole form inspired
by analogy with the nucleon elastic form factors. Correspondingly,
we consider here two parametrizations of the two--gluon form factor:
\beq
F_{2g} (x, t|Q^2) \;\; = \;\;
\left\{ \begin{array}{l}
\displaystyle
\exp (B_g t/2) ,
\\[2ex]
\displaystyle
(1 - t/m_g^2)^{-2} ,
\end{array}
\right.
\label{twogl_exp_dip}
\eeq
where the parameters $B_g$ and $m_g$ are functions of $x$ and $Q^2$.
The two parametrizations give very similar results if the functions
are matched at $|t| = 0.5 \, \mbox{GeV}^2$, where they are best
constrained by present data (see Fig.~3 of Ref.~\cite{Frankfurt:2006jp});
this corresponds to \cite{Frankfurt:2010ea}
\beq
B_g \;\; = \;\; 3.24/m_g^2 .
\label{dip_exp}
\eeq
The analysis of the HERA exclusive data leads to
\begin{eqnarray}
B_g (x) = B_{g0} \; + \; 2 \alpha'_g \; \ln (x_0/x) ,
\label{bg_param}
\end{eqnarray}
where $x_0 = 0.0012,
B_{g0} = 4.1 \; ({}^{+0.3}_{-0.5}) \; \mbox{GeV}^{-2}, \alpha'_g = 0.140 \; ({}^{+0.08}_{-0.08}) \; \mbox{GeV}^{-2}$ for $Q_0^2\sim $ 3 GeV$^2$.  For fixed $x$, $B(x,Q^2)$ slowly decreases with increase of $Q^2$  due to the DGLAP evolution
\cite{FSW}. The uncertainties in parentheses represent a rough estimate
based on the range of values spanned by the H1 and ZEUS fits,
with statistical and systematic uncertainties added linearly. This estimate does not include possible contributions to  $\alpha'_g$ due to the contribution of the  large size configurations in the vector mesons and changes in the evolution equation at
$-t$ comparable to the intrinsic scale.  Correcting for these effects may lead to a reduction of $\alpha'_g$ and hence to a slower increase of
the area occupied by gluons with decrease of $x$.

\end{document}